\documentclass[a4paper,11pt]{article}
\usepackage{jheppub} 
\usepackage{multirow}
\usepackage{lineno}

\arxivnumber{xxxx.xxxxx} 
\title{\boldmath Antineutron reconstruction in electromagnetic calorimeters with mixed-representation learning}

\author[a,1]{Yangu Li,}
\author[b,c,1]{Hongtian Yu,}
\author[d,c]{Yuyang Huang,}
\author[e,c]{Zhi Cao,}
\author[f]{Yunxuan Song,}
\author[b]{Yunfan Liu,}
\author[g]{Yajun Mao,}
\author[a]{YangHeng Zheng,}
\author[a,2]{Xiao-Rui Lyu,}
\author[b,2]{and Qixiang Ye}
\affiliation[a]{School of Physical Sciences, University of Chinese Academy of Sciences,\\No.1 Yanqihu East Rd, Beijing 101408, China}
\affiliation[b]{School of Electronic, Electrical and Communication Engineering, University of Chinese Academy of Sciences,\\No.1 Yanqihu East Rd, Beijing 101408, China}
\affiliation[c]{Faculty of AI for Natural Sciences, Zhongguancun Academy,\\No.17 Daniufang 2nd Ring Rd, Beijing 100094, China}
\affiliation[d]{School of Instrument Science and Engineering, Harbin Institute of Technology,\\No.92 West Dazhi Street, Harbin 150001, China}
\affiliation[e]{School of Computer Science and Engineering, Beihang University,\\No.37 Colleage Road, Beijing 100191, China}
\affiliation[f]{School of Nuclear Science and Technology, University of Chinese Academy of Sciences,\\No.1 Yanqihu East Rd, Beijing 101408, China}
\affiliation[g]{School of Physics, Peking university,\\No.209 Chengfu Road, Beijing 100871, China}
\note[1]{These authors contributed equally to this work.}
\note[2]{Corresponding author.}
\emailAdd{liyangu@ucas.ac.cn}
\emailAdd{yuhongtian17@mails.ucas.ac.cn}
\emailAdd{24b901043@stu.hit.edu.cn}
\emailAdd{zhicao@buaa.edu.cn}
\emailAdd{yunxuan.song@cern.ch}
\emailAdd{liuyunfan@ucas.ac.cn}
\emailAdd{maoyj@pku.edu.cn}
\emailAdd{zhengyh@ucas.ac.cn}
\emailAdd{xiaorui@ucas.ac.cn}
\emailAdd{qxye@ucas.ac.cn}

\compress
\abstract{
    A long-standing bottleneck in GeV-scale accelerator experiments lies in reconstructing long-lived neutral hadrons in conventional electromagnetic calorimeters (ECALs), where hadron--nucleus interactions fall outside the detector's native response regime.
    In this paper, we develop a physics-inspired representation approach for antineutron reconstruction using a large corpus of real collision data. Motivated by two distinct energy deposition patterns from the penetrating high energy antineutrons in ECALs, we propose a Mixed-representation Calorimetric Network (MrCAL) that integrates complementary visual and sequential representation branches within a unified object-detection architecture. This architecture jointly predicts particle identity, momentum direction, and momentum magnitude.
    Our approach improves the precision of antineutron momentum-direction reconstruction by up to 96\% and, for the first time, enables direct measurement of momentum magnitude solely from ECAL readouts, achieving a momentum resolution of approximately 17\% at 1~${\rm GeV}/c$. 
    The model maintains robust performance through comprehensive generalization tests spanning a wide variety of physics processes and background environments.
    This work unlocks a novel measurement capability for legacy ECAL systems at large experimental facilities, broadening their scientific scope via innovative final-state neutral-hadron detection.
}

\begin{document}
\maketitle
\flushbottom

\section{Introduction}
A central task in experimental high-energy physics is particle reconstruction~\cite{Fruhwirth:2020zbo,YUSSUP2026113377}: the inference of particle identities and kinematics from detector responses. Charged particles are primarily reconstructed with tracking systems, photons and electrons with electromagnetic calorimeters (ECALs), and muons with penetration depth in dedicated muon systems. Long-lived neutral hadrons, including neutrons, antineutrons, and $K_L^0$ mesons, are much difficult to reconstruct, as they produce no ionization trajectories in tracking system. In many GeV-scale experiments~\cite{BaBar:2001yhh,CLEO:1991qyy,BESIII:2009fln,Belle-II:2018jsg,Achasov:2023gey} their reconstruction relies solely on hadronic energy depositions recorded in calorimeter subsystems, most notably ECAL.

Reconstruction of long-lived neutral hadron based on ECAL is intrinsically challenging. ECAL hardware and software are optimized to capture electromagnetic showers arising from bremsstrahlung and electron–positron pair production, whereas neutral hadrons undergo through nuclear scattering, spallation and absorption processes, as well as annihilation process occurring for antimatter counterparts~\cite{Pietropaolo:2020frm}. The geometric layout and material composition of ECAL are, therefore, ill-suited to accommodate these intricate hadronic interactions,
hampering both the efficient initiation and complete containment of hadronic showers. Consequently, neutral hadrons yield sparse, fragmented and non-local energy deposition signatures in ECAL, rendering conventional clustering algorithms~\cite{He:2011zzd} incapable of delivering robust reconstructions of the incoming particle’s direction and momentum.

This limitation has direct consequences for GeV-scale particle physics. A neutral hadron presented in the reaction final states can turn an otherwise exclusive measurement into a partial reconstruction problem, broaden invariant mass peaks, degrade background suppression performance, and hinder extraction of angular and spin observables. Such final-state topologies frequently arise in studies covering hyperon and heavy-flavor hadron decays, baryon spectroscopy, exotic hadron searches, and neutron internal structure measurements~\cite{BESIII:2022rgl,BESIII:2015jmz,BESIII:2021tbq,BESIII:2024mgg}. Enhancing neutral hadron reconstruction capability is therefore a prerequisite for expanding the physics reach of these physics programs.

More broadly, this reconstruction task exemplifies a general category of inverse problems encountered across scientific instrumentation, in which the informative signals are encoded in detector responses lying outside the device's native operational regime. Analogous challenges arise, for example, in multiwavelength astronomy, where ultraviolet, visible, and near-infrared observations must be combined to recover information that no single band can fully capture~\cite{Scoville:2006vq,10.1111/j.1365-2966.2008.13535.x}. In this context, neutral hadron reconstruction becomes an instance of representation learning under physical constraints, for which modern machine learning (ML) techniques offer a natural and principled solution.

In recent years, ML has been extensively explored for particle reconstruction in calorimeters, with a wide range of data representations and model architectures proposed~\cite{Paganini:2017dwg,Chekalina:2018hxi,Belayneh:2019vyx,Alimena:2020web,Akchurin:2021afn,Polson:2021kvr,Khattak:2021ndw,Krause:2021ilc,Bieringer:2022cbs,Rogachev:2022hjg,NA62:2023wzm,Song:2023ceh,Charan:2023ldg,Simkina:2023ztj,Dubinski:2023fsy,Simsek:2024zhj,Hashmani:2024ykk,Favaro:2024rle,Liu:2024kvv,Qasim:2022rww,Belle-II:2023cal,CMSHGCAL:2024esz,Kobylianskii:2024ijw,Akchurin:2024ffj,Acosta:2023nuw,Buhmann:2023bwk}. These methods have proved impressively gains in reconstruction accuracy and efficiency~\cite{Belayneh:2019vyx,Alimena:2020web,Akchurin:2021afn,Polson:2021kvr,NA62:2023wzm,Song:2023ceh,Charan:2023ldg,Belle-II:2023cal,Simkina:2023ztj,CMSHGCAL:2024esz,Akchurin:2024ffj}, background and pileup mitigation~\cite{Qasim:2022rww,Hashmani:2024ykk,Schnake:2024mip,Dimitrova:2025mbl}, and shower simulation~\cite{Paganini:2017dwg,Chekalina:2018hxi,Belayneh:2019vyx,Khattak:2021ndw,Krause:2021ilc,Bieringer:2022cbs,Rogachev:2022hjg,Dubinski:2023fsy,Buhmann:2023bwk,Favaro:2024rle,Krause:2024avx,Simsek:2024zhj,Liu:2024kvv,Kobylianskii:2024ijw,Schnake:2024mip}. Despite this progress, most existing studies focus on reconstruction problems within the native operating regime of the calorimeter. The reconstruction of neutral hadrons in ECALs has received little systematic attention.

Moreover, only a few proposed ML methods for calorimeter reconstruction have been deployed in ongoing experiments~\cite{NA62:2023wzm,CMSHGCAL:2024esz}. Practical adoption is often limited by the quality and realism of the available datasets, particularly by discrepancies between simulation and real data. It is also constrained by concerns regarding robustness, stability under evolving detector conditions, and interpretability in downstream physics analyses. These limitations highlight the need for reconstruction methods that are not only effective in controlled benchmarks, but also generalizable under realistic experimental conditions to deliver unbiased, precise physics measurements.

\begin{figure}[ht]
    \centering
    \includegraphics[width=\linewidth]{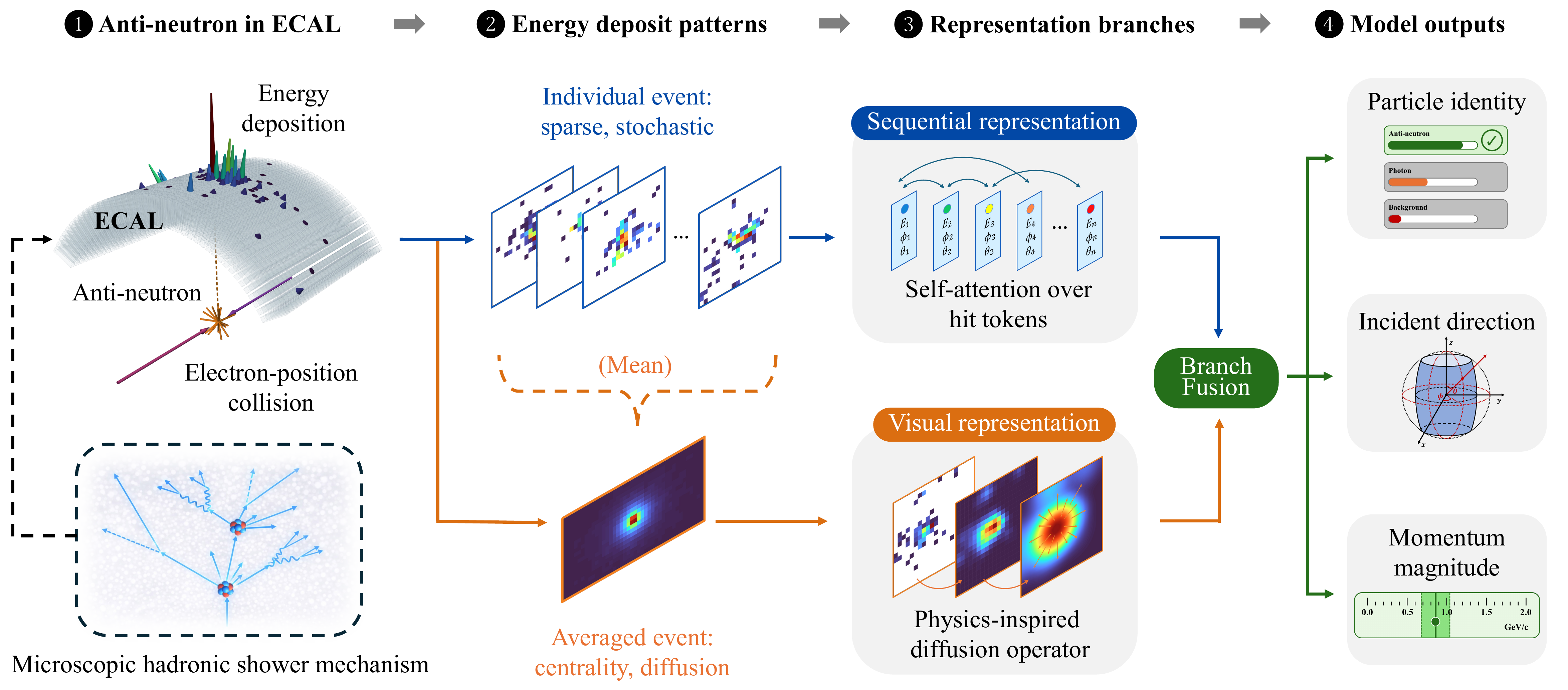}
    \caption{Overview of mixed-representation calorimetric network (MrCAL). Hadronic interactions of antineutrons in an ECAL produce two distinct regularities in the deposited energy: sparse, stochastic and non-local patterns in individual events, and a reproducible centrality-and-diffusion prior in averaged events. MrCAL exploits these patterns by combining a sequential branch that aggregates dispersed hit tokens with a visual branch that encodes the averaged spatial prior through a diffusion operator. The fused framework jointly reconstructs particle identity, incident direction, and momentum magnitude from ECAL information alone.}
    \label{fig:intro}
\end{figure}

In this work, we develop a physics-inspired representation approach for neutral hadron reconstruction in ECALs, with an initial focus on antineutrons. Antineutrons can annihilate in the ECAL material, producing larger and more informative energy deposits than neutrons, and thus providing a natural starting point for learning-based reconstruction. By analyzing antineutron energy deposition patterns in the ECAL, we identify two dominant empirical regularities, as illustrated in Fig.~\ref{fig:intro}. For an individual event, the response is sparse, highly stochastic and spatially non-local, with informative deposits often scattered across disconnected regions. When averaged over many events, however, the same response reveals a reproducible spatial prior, a central concentration around the incident point together with an approximately radial, diffusion-like attenuation profile. 

These observations lead us to formulate antineutron reconstruction as a problem of mixed-representation learning. We design complementary branches under a unified object-detection~\cite{voc, mscoco} framework: a sequential branch that models long-range dependencies and non-local correlations among dispersed deposits, and a visual branch that encodes the spatial priors of centrality and diffusion. Their integration yields a mixed-representation calorimetric network (MrCAL), in which particle type, momentum direction, and momentum magnitude are inferred jointly from ECAL information alone.

Using a large corpus of real collision data collected in the representative BESIII experiment~\cite{BESIII:2009fln}, we perform a comprehensive evaluation of MrCAL under realistic detector conditions. The angular resolution of antineutron direction is improved by up to 96\%, and, for the first time, a reliable prediction of the momentum magnitude is achieved, reaching a resolution of approximately 17\% at 1 $\mathrm{GeV}/c$. Extensive tests across broad kinematic and background conditions demonstrate stable performance and strong generalization, indicating feasibility for deployment in real experiments and broad potential for physics analyses.

\section{Model architecture}
\subsection{Visual representation branch}
To capture the statistically stable component of antineutron responses in ECALs, we construct a visual representation that maps calorimeter cell readouts onto a two-dimensional image and formulates reconstruction as an object detection problem. This representation is designed to exploit the averaged centrality-and-diffusion pattern of hadronic energy deposits, which is less apparent in any single sparse event but becomes reproducible at the population level, as illustrated in Fig.~\ref{fig:intro}. In practice, it is implemented as a network branch adapted from modern visual recognition pipelines.

The visual branch incorporates three key design elements. First, a physics-inspired diffusion operator~\cite{vheat2025} encodes the averaged spatial prior. By propagating sparse local activations in the frequency domain through a radially symmetric diffusion kernel, the branch builds features that are sensitive to broad shower morphology rather than isolated high-energy cells. Second, as the reconstruction target is an incident point on the ECAL surface rather than a conventional object extent~\cite{retinanet}, point regression is embedded to an object-detection formulation by assigning pseudo bounding boxes around the labeled incident positions. These pseudo boxes provide contextual scale for detection supervision while preserving the physical interpretation of the target. Third, the multi-scale detection architecture allows local spatial features to support incident-direction inference while broader contextual features contribute to momentum-magnitude prediction. Technical details are given in Appendix~\ref{sec:app_vis}.

\subsection{Sequential representation branch}
To capture event-by-event fluctuations of antineutron responses in the ECAL, we construct a sequential representation in which each event is encoded as a token sequence of active calorimeter cells. The representation is implemented with a transformer-based~\cite{attention} sequential network branch, as sketched in Fig.~\ref{fig:intro}. Complementary to the visual branch, the sequential representation preserves the sparse active cells in individual events and explicitly models non-local correlations among dispersed deposits. It is therefore more information-efficient than the visual representation in the regime where only a small fraction of ECAL cells are active.

Operationally, each active cell is treated as a token carrying its deposited energy and angular coordinates. Events with different hit multiplicities are mapped to a fixed sequence length through repetition and padding. To avoid imposing an artificial ordering on the unordered hit set, tokens are randomly permuted within each repeated segment, while cyclic shifts of the azimuthal coordinate emulate the rotational symmetry of the ECAL surface. The deposited energy and angular coordinates are discretized and embedded through learnable lookup tables, and a learnable momentum token is appended to aggregate global-level information. Self-attention, controlled by an explicit mask, aggregates valid sparse deposits while suppressing padding artifacts. The incident point is inferred through token-level proposals that form an internal voting process, whereas the momentum magnitude is regressed from the momentum token. A position-cloze pretraining strategy further improves the representation quality: selected medium-to-high-energy deposits have their angular coordinates masked, and the model predicts these coordinates from energy prompts and the surrounding hit context, thereby learning the spatial pattern without using incident labels. Technical details are given in Appendix~\ref{sec:app_seq}.

\subsection{Mixed representation of two branches}
The visual and sequential branches encode complementary aspects of antineutron energy-deposition patterns in the ECAL, which can be either deployed independently or combined to a mixed-representation model. In MrCAL, the two branches process the same ECAL information through different representations and retain branch-specific prediction pathways under a unified interface, allowing their distinct inductive biases to contribute jointly to reconstruction.

During training, the mixed-representation model is optimized by combining the branch-specific losses over the same reconstruction targets. During inference, the two branches first generate candidate detections independently, including predicted incident locations, pseudo boxes, confidence scores and associated kinematic outputs. These candidates are then calibrated and reconciled according to class consistency, angular agreement and confidence. Consistent branch-level predictions are aggregated into refined kinematic predictions, ambiguous pairs are resolved by selecting the higher-confidence branch, and unmatched high-confidence candidates are retained. Finally, MrCAL converts the candidates into a unified set of detections in which particle type, incident direction, and momentum magnitude are reconstructed. Technical details are given in Appendix~\ref{sec:app_mix}.

\section{Real-data benchmark}\label{sec:data}
A central ingredient of this study is a large real-data benchmark of antineutrons in the BESIII ECAL~\cite{BESIII:2009fln}. While the \textsc{Geant4} toolkit~\cite{GEANT4:2002zbu} is capable to simulate particle--material interactions with high precision, significant discrepancies, up to the $20\%$ level, have been observed between data and simulation for neutral-hadron responses in sub-GeV region~\cite{LIU2022166672}. These discrepancies originate in part from insufficient experimental input on the relevant interaction cross sections~\cite{Gunderson:1979yi,BROOKHAVEN-HOUSTON-PENNSYLVANIASTATE-RICE:1987vhf,OBELIX:2000kga}, as well as from imperfect estimates of beam backgrounds and detector noise. A reconstruction method trained and validated only on idealized simulation could therefore learn simulation-specific regularities rather than true antineutron signatures, highlighting the need of a real-data benchmark.

The dataset is assembled from three representative BESIII collision samples: $8.772\times10^9$ $J/\psi$ decays at a center-of-mass energy of $\sqrt{s}=3.097\,\mathrm{GeV}$~\cite{BESIII:2021cxx}, $2.259\times10^9$ $\psi(2S)$ decays at $\sqrt{s}=3.686\,\mathrm{GeV}$~\cite{BESIII:2024lks}, and $17.3\,\mathrm{fb}^{-1}$ of $e^+e^-$ annihilation data at $\sqrt{s}=3.773\,\mathrm{GeV}$~\cite{BESIII:2024lbn}. These samples provide a broad set of exclusive baryonic final states containing one antineutron, including topologies of $p\bar n\pi^-$, $p\bar n\pi^-\pi^+\pi^-$, $\Lambda\bar\Lambda$, $\Sigma^+\bar\Sigma^-$, $pK^-\bar\Lambda$, $\Lambda\bar\Sigma^\mp\pi^\pm$, $\Lambda\bar\Lambda\pi^+\pi^-$, and $\Lambda\bar n K^0_S$, with the antineutron produced either directly or through antihyperon decays. For the $\psi(2S)$ sample, both its direct decays and cascade decays through $\psi(2S)\to\pi^+\pi^-J/\psi$ are included. 

In total, we investigate 36 decay channels with 5,243,268 selected events, and define non-overlapping channel sets for the benchmarks, as detailed in Table~\ref{tab:dst}. For model training and nominal performance evaluation in Sec.~\ref{sec:performance}, we construct a momentum-balanced antineutron sample through an acceptance--rejection procedure, yielding 1.1 million events with an approximately uniform momentum distribution over $[0.05,\,1.5]\,\mathrm{GeV}/c$. Of these events, 1.0 million are used for training and validation, and 0.1 million are reserved for testing. For the generalization study in Sec~\ref{sec:generalization}, we evaluate across 28 channels, including a held-out subset of 13 channels that is fully excluded from training.

\begin{table}[ht]
\footnotesize
\centering
\caption{\label{tab:dst}
Benchmarking channels used for model training and nominal performance evaluation (P), generalization tests in the single-particle setting (S), and generalization tests in the multi-particle setting (M).}
\begin{tabular}{c|c|c|c|c|c|c|c|c|c|c}
\hline\hline
& & $p\bar{n}\pi^-$ & $p\bar{n}\pi^-\pi^+\pi^-$ & $\Lambda\bar{\Lambda}$ & $\Sigma^+\bar{\Sigma}^-$ & $pK^-\bar{\Lambda}$ & $\Lambda\bar{\Sigma}^-\pi^+$ & $\Lambda\bar{\Sigma}^+\pi^-$ & $\Lambda\bar{\Lambda}\pi^+\pi^-$ & $\Lambda\bar{n}K^0_S$ \\ 
\hline
\multirow{3}{6em}{\centering $J/\psi$\\decays} & P & \checkmark & \checkmark & \checkmark & \checkmark & \checkmark & \checkmark & \checkmark & \checkmark & \checkmark \\ 
& S & \checkmark & \checkmark & \checkmark & \checkmark & \checkmark & \checkmark & \checkmark & \checkmark & \checkmark \\
& M & - & - & \checkmark & \checkmark & \checkmark & - & - & \checkmark & - \\
\hline
\multirow{3}{6em}{\centering $\psi(2S)$\\decays} & P & \checkmark & \checkmark & - & - & - & - & - & \checkmark & - \\ 
& S & - & \checkmark & \checkmark & \checkmark & - & - & - & \checkmark & \checkmark \\
& M & - & - & \checkmark & \checkmark & - & - & - & \checkmark & - \\
\hline
\multirow{3}{6em}{\centering $\psi(2S)\to\pi^+\pi^-J/\psi$ decays} & P & - & \checkmark & \checkmark & \checkmark & \checkmark & - & - & - & - \\ 
& S & \checkmark & \checkmark & \checkmark & \checkmark & \checkmark & \checkmark & \checkmark & \checkmark & \checkmark \\
& M & - & - & \checkmark & \checkmark & \checkmark & - & - & \checkmark & - \\
\hline
\multirow{3}{6em}{\centering $e^+e^-$ at $3.773\,{\rm GeV}$} & P & \checkmark & \checkmark & - & - & - & - & - & - & - \\ 
& S & - & - & - &- & \checkmark & \checkmark & \checkmark & \checkmark & \checkmark \\
& M & - & - & - & - & \checkmark & - & - & \checkmark & - \\
\hline
\end{tabular}
\end{table}

For each selected event, the model input consists of the deposited energy and angular coordinates of all ECAL hits, parameterized by the azimuthal angle $\phi$ and polar angle $\theta$. To suppress contamination from other reconstructed final-state particles, ECAL hits within a $30^\circ$ cone around their extrapolated incident points are masked. The antineutron momentum direction and magnitude used as supervision labels are obtained with a recoil technique against the reconstructed four-momenta of the remaining final-state particles. These recoil constraints are often unavailable in general physics studies, and therefore does not diminish the significance of this work.

A more complicated and realistic scenario is considered in this work where multiple neutral particles exist in the decay final state. Since sufficiently pure and abundant real-data samples are limited for this case, we construct a hybrid dataset by merging one real-data antineutron pattern with two photon patterns from Monte Carlo simulation. The photons are generated using single particle gun events with isotropic directions and momenta uniformly sampled in $[0,\,1.5]\,{\rm GeV}/c$. This strategy leverages the comparatively high fidelity of Geant4 for photon interactions while preserving realistic beam and detector backgrounds already present in the real-data antineutron events. The hybrid datasets are used solely for ML training and for the nominal performance benchmarks. In the multi-particle generalization study, we instead evaluate on channels in Table~\ref{tab:dst} that naturally contain an antineutron accompanied by two photons, retaining the original hit-level information without hybrid mixing.

\section{Reconstruction performance}\label{sec:performance}
We evaluate the visual branch, the sequential branch and the fused MrCAL model using the BESIII antineutron benchmark described in Sec.~\ref{sec:data}. Reconstruction performance is assessed through three quantities required by downstream physics analyses: particle identity, incident direction and momentum magnitude. Directional precision is quantified by the mean angular bias (mAB), and momentum precision by the mean absolute error (mAE) and mean relative error (mRE). The model outputs include a confidence score, enabling an explicit efficiency--precision tradeoff through score-threshold scans.

\begin{figure}[b]
    \centering
    \includegraphics[width=\linewidth]{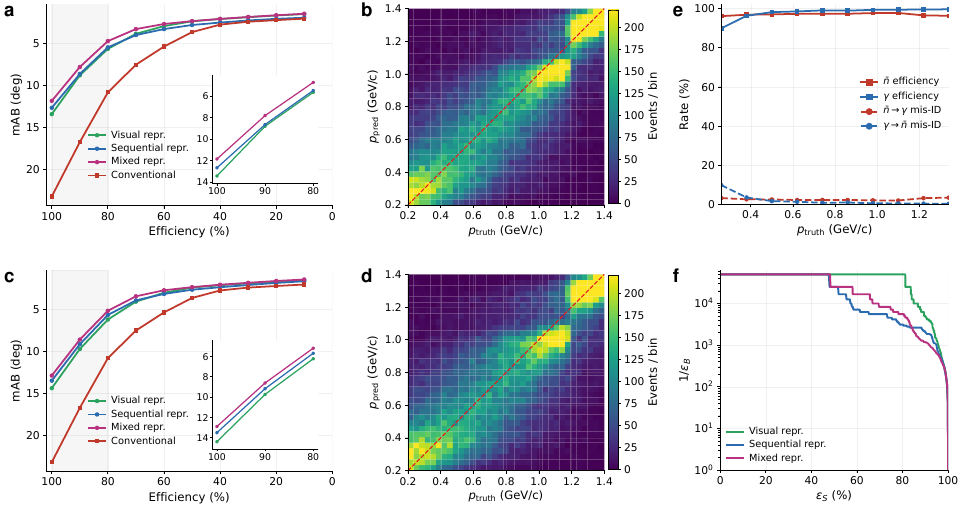}
    \caption{Reconstruction performance. (a) Efficiency dependence of the mAB in the single-particle setting. (b) Correlation between true and predicted momentum magnitudes in the single-particle setting. (c) Efficiency dependence of the mAB in the multi-particle setting. (d) Correlation between true and predicted momentum magnitudes in the multi-particle setting. (e) Antineutron and photon identification efficiencies and misidentification rates as functions of the true momentum magnitude. (f) Signal efficiency $\epsilon_{S}$ versus background rejection $1/\epsilon_{B}$ for event-level classification.}
    \label{fig:eval}
\end{figure}

Evaluation begins with the most idealized scenario, namely the single-particle setting, where each event is known to contain a single antineutron and the detected object with the highest confidence score is taken as the reconstruction output. A conventional clustering method~\cite{He:2011zzd} widely adopted at BESIII is used as the baseline, in which the incident direction of the antineutron is approximated by the center of the most energetic ECAL cluster. All three learning-based configurations improve direction measurement relative to this conventional method. At full efficiency, the mAB is reduced from 23.2$^\circ$ for the conventional method to 13.4$^\circ$ for the visual branch, 12.7$^\circ$ for the sequential branch and 11.8$^\circ$ for MrCAL, as shown in Fig.~\ref{fig:eval}(a). Using $1/{\rm mAB}$ as the directional precision, these values correspond to gains of 73\%, 82\%, and 96\%, respectively. At low efficiency, the performance of all methods approaches a similar precision limit, set primarily by the finite angular granularity of ECAL cells and the residual uncertainty of the truth labels. The angular precision also exhibits a clear positive dependence on antineutron momentum magnitude, as shown in Fig.~\ref{fig:supp_eval_single}(b) in Appendix~\ref{sec:suppfig}, consistent with the expectation that higher-momentum antineutrons tend to leave more informative signatures in the ECAL.

Crucially, the same approach also enables direct regression of the antineutron momentum magnitude from ECAL information alone. According to Fig.~\ref{fig:eval}(b), the predicted momentum is approximately unbiased and nearly linear with the true momentum over the range 0.2--1.4~${\rm GeV}/c$. At 1~${\rm GeV}/c$, the momentum resolution reaches 17.8\% for the visual branch, 16.7\% for the sequential branch and 17.6\% for MrCAL. The achieved resolution is limited mainly by intrinsic energy leakage and random fluctuations in hadronic shower development, together with the limited containment of ECAL. Nevertheless, it is already better than what is typically obtained with dedicated hadronic calorimeters in the sub-GeV regime, including LHC-era systems where resolutions above 50\% are common at these energies~\cite{Cavallari_2011}. The efficiency and momentum dependencies of the mRE and mAE are shown in Fig.~\ref{fig:supp_eval_single} in Appendix~\ref{sec:suppfig}.

We further test the model in a more complicated and realistic multi-particle setting containing one antineutron accompanied by photons. The unified reconstruction pipeline maintains stable antineutron performance in the presence of additional ECAL activity. The mAB at full efficiency is 14.3$^\circ$, 13.5$^\circ$ and 12.9$^\circ$ for the visual branch, the sequential branch and MrCAL, respectively; the corresponding momentum resolutions at 1~${\rm GeV}/c$ remain close to the single-particle case at 18.7\%, 18.2\% and 18.2\%, as shown in Fig.~\ref{fig:eval}(c,d). The same pipeline also preserves photon reconstruction at a satisfied performance level expected from an ECAL, as shown in Fig.~\ref{fig:supp_eval_multi_g} in Appendix~\ref{sec:suppfig}. These performances rely on robust object-level particle identification, with antineutron and photon misidentification rates of only 2.6\% and 2.0\%, respectively, according to Fig.~\ref{fig:eval}(e). 

In addition to object-level reconstruction, our approach provides an auxiliary event-level classification output that can be used as a physics analysis tool for isolating rare signal events from massive backgrounds~\cite{BESIII:2022rgl,BESIII:2015jmz,BESIII:2021tbq,BESIII:2024mgg}. In a binary discrimination task between single-particle signal and multi-particle background events, MrCAL reduces the background by two orders of magnitude while retaining 99.7\% of the signals, and by three orders of magnitude while retaining 91.7\% of the signals, as shown in Fig.~\ref{fig:eval}(f). The full performance evaluation including branch-wise reconstruction metrics and photon performance is provided in Appendix~\ref{sec:suppfig}.

\section{Generalization tests}\label{sec:generalization}
Performance gains are meaningful for experimental application only if they persist across the diverse conditions encountered in real physics interactions, including kinematic regions, event topologies, and background compositions. Antineutron signatures are particularly sensitive to such changes because the informative energy deposits are sparse and can be easily obscured by accompanying particles or random noise. We therefore stress-test the three learning-based configurations using 28 independent benchmarking channels, including 13 channels held out entirely from training, as listed in Table~\ref{tab:dst}. 

\begin{figure}[htbp]
    \centering
    \includegraphics[width=\linewidth]{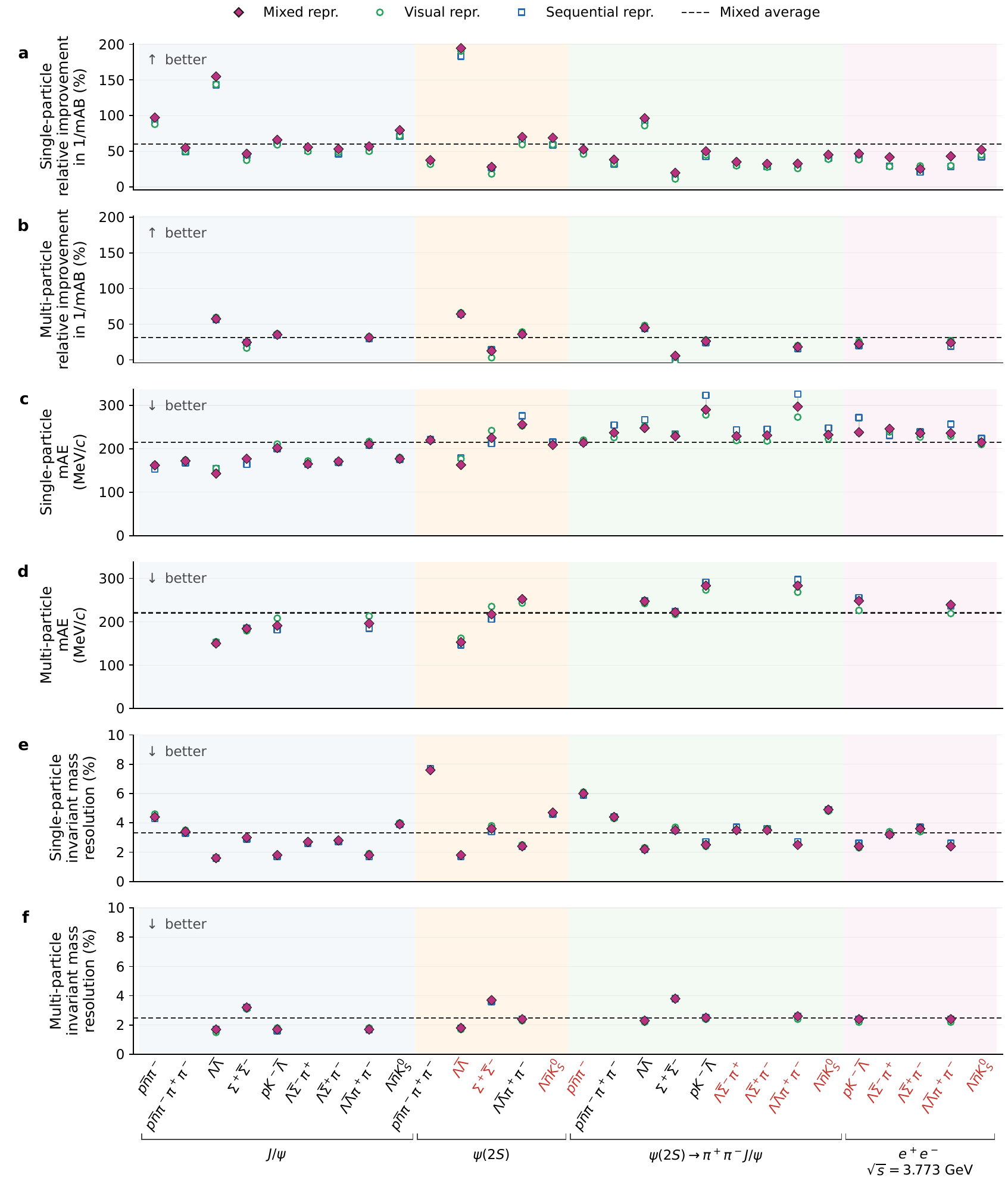}
    \caption{Generalization tests. (a) Relative improvement in directional precision across benchmarking channels in the single-particle setting. (b) The same metric in the multi-particle setting. (c) Absolute error of momentum-magnitude regression across benchmarking channels in the single-particle setting. (d) The same metric in the multi-particle setting. (e) Invariant mass resolutions of short-lived mother particles reconstructed using MrCAL-predicted antineutron four-momentum in the single-particle setting. (f) The same metric in the multi-particle setting. Benchmarking channels held out from training are highlighted in red.}
    \label{fig:gene}
\end{figure}

The improvement in antineutron direction reconstruction is consistently retained across the benchmark. In the single-particle setting, MrCAL improves the channel-specific directional precision by 194.5\% in maximum, by 19.8\% in minimum, and by 59.8\% in average relative to the conventional method, as shown in Fig.~\ref{fig:gene}(a). In the more challenging multi-particle setting, the average gain remains 31.0\% according to Fig.~\ref{fig:gene}(b). Momentum-magnitude regression is also stable across benchmarking channels, with the channel-averaged mAE remain 214~${\rm MeV}/c$ in single-particle setting and 220~${\rm MeV}/c$ in multi-particle setting, as shown in Fig.~\ref{fig:gene}(c,d). These results show that the models learn transferable features of antineutron responses in the ECAL rather than exploiting channel-specific regularities.

We further validate the approach in a representative physics analysis task by reconstructing short-lived particles that decay to antineutrons, including $\Lambda$ and $\Sigma^\pm$ hyperons as well as $J/\psi$ and $\psi(2S)$ charmonia~\cite{ParticleDataGroup:2024cfk}. The model-predicted antineutron four-momentum is combined with standard reconstructions of the accompanying final-state particles to form the mother particle. Clear invariant-mass peaks are recovered across the benchmark channels, with unbiased central positions and controlled widths, as shown in Fig.~\ref{fig:supp_gene_invm_single} and Fig.~\ref{fig:supp_gene_invm_multi} in Appendix~\ref{sec:suppfig}. For MrCAL, the average invariant-mass resolution is 3.3\% in the single-particle setting and 2.5\% in the multi-particle setting, as shown in Fig.~\ref{fig:gene}(e,f). When model-predicted photon four-momenta are also incorporated, the average resolution becomes 3.1\% in Fig.~\ref{fig:supp_gene_multi_g}(d) in Appendix~\ref{sec:suppfig}. Overall, these results show that the reconstruction ability can propagate coherently to physics observables.

\section{Conclusions and outlook}
This work shows that a persistent limitation in GeV-scale accelerator experiments, the reconstruction of long-lived neutral hadrons in ECALs, can be substantially mitigated through physics-inspired representation learning. The core of this approach is the mixed-representation calorimetric network (MrCAL), which reframes neutral hadron reconstruction as an inverse problem outside the detector’s native response regime. MrCAL recovers kinematic information from energy deposits that are sparse, fragmented and non-local. Quantitatively, it improves the antineutron momentum-direction resolution by up to 96\% and enables direct regression of the momentum magnitude from ECAL information alone, reaching a resolution of about 17\% at 1~${\rm GeV}/c$. These gains persist across broad kinematic regions, held-out channels, and complex neutral environments, demonstrating good robustness.

A comparison of the visual and sequential representation branches further clarifies the value of mixed representation. For momentum-direction reconstruction, the two branches show similar standalone performance, whereas their fusion in MrCAL yields consistently better precision, indicating that they capture complementary aspects of the task. For momentum-magnitude reconstruction, the two branches exhibit a clearer trade-off: the sequential branch performs better in the nominal training domain, whereas the visual branch generalizes more robustly beyond it. In this case, MrCAL reaches a balanced performance between the two, combining strong in-domain accuracy with out-of-domain stability. Taken together, the convergence of distinct approaches and the diminishing returns from increased architectural complexity suggest that the current performance may already be approaching a regime set largely by detector design, labeling bias and intrinsic fluctuation of hadronic shower in the ECAL.

Our approach restores a basic detector capability that has remained structurally limited in many particle physics experiments~\cite{BaBar:2001yhh,CLEO:1991qyy,BESIII:2009fln,Belle-II:2018jsg,Achasov:2023gey}, and the implications for running experiments like BESIII are immediate and practical. Because the approach is purely algorithmic, it can be deployed on existing data without hardware upgrades or interruptions to operation. This effectively expands the performance envelope and scientific reach of an existing experiment at negligible cost. Improved antineutron reconstruction directly benefits physics analyses in which neutral baryons play a central role~\cite{BESIII:2022rgl,BESIII:2015jmz,BESIII:2021tbq,BESIII:2024mgg}. Some concrete examples include the studies of $CP$ violation and quantum number violation in baryons, which probe fundamental sources of matter–antimatter asymmetry in the universe~\cite{Sakharov:1967dj}. Many baryons of interest decay to final states containing (anti)neutrons~\cite{ParticleDataGroup:2024cfk}, for which limited reconstruction capability becomes a dominant experimental bottleneck. By lifting this limitation, MrCAL can enable more precise measurements and open access to decay channels that were previously constrained by detector performance.

Beyond the specific case of antineutrons at BESIII, the underlying formulation naturally extends to a broader class of particles and experimental settings. Other long-lived neutral hadrons, most notably the $K^0_L$ meson, face the same mismatch between hadronic interaction mechanisms and ECAL design. Related challenges also arise in other GeV-scale accelerator experiments, including the finished BaBar~\cite{BaBar:2001yhh}, Belle~\cite{Belle:2000cnh} and KLOE~\cite{Bossi:2008aa}, the ongoing Belle~II~\cite{Belle-II:2018jsg}, JLab experiments~\cite{CLAS:2003umf} and sPHENIX~\cite{Campbell:2016cea}, as well as the proposed STCF~\cite{Achasov:2023gey}, PANDA~\cite{Belias:2023lkk} and HIAF~\cite{Anderle:2021wcy}, where neutral hadron reconstruction in ECALs remains constrained. For TeV-scale experiments equipped with dedicated hadronic calorimeters, such as the LHC experiments~\cite{CMS:2008xjf,ATLAS:2008xda,ALICE:2008ngc,LHCb:2008vvz}, the proposed CEPC~\cite{CEPCStudyGroup:2018ghi} and FCC~\cite{FCC:2018byv}, ML strategies could further improve their reconstruction performances. More broadly, similar reconstruction tasks also appear in high-energy neutrino and astroparticle experiments, including JUNO~\cite{JUNO:2021vlw} and LHAASO~\cite{LHAASO:2019qtb}, where sparse and stochastic detector responses encode essential physical information. Across these cases, the proposed MrCAL framework provides a directly transferable solution.

The findings also point toward a broader paradigm for particle detector design. As ML-based reconstruction gains greater predictive power, the classical partition between electromagnetic and hadronic calorimeter measurements may become less rigid. Future detectors will increasingly be optimized not only to support native analytical reconstruction, but also to deliver higher-dimensional, higher-granularity, and well-calibrated raw information tailored for ML-driven inference. In this view, the effective functionality of a detector can be expanded via co-design of hardware architectures and the representational capacity of state-of-the-art ML.  Antineutron reconstruction solely within an ECAL serves as a concrete demonstration of this design paradigm shift.

\section*{Acknowledgements}
The authors thank the BESIII Collaboration for providing the labeled antineutron samples, and thank the IHEP computing center and the supercomputing center of Lanzhou university for their strong support. 
This work is supported in part by National Key R\&D Program of China under Contract No. 2025YFA1613900; CAS Project for Young Scientists in Basic Research under Contract No. YSBR-117; National Natural Science Foundation of China (NSFC) under Contracts Nos. 62225208 and 12221005; Zhongguancun Academy under Contract No. 02012412.

\section*{Data availability}
The data that support the findings of this article are not publicly available upon publication because it is not technically feasible and/or the cost of preparing, depositing, and hosting the data would be prohibitive within the terms of this research project. Access to the data is governed by the BESIII Collaboration data policy and handled according to the relevant collaboration approval procedures. The code used to analyze the data is available from Ref.~\cite{github}.

\appendix
\section{Design details of model architecture}
\subsection{Visual representation branch}\label{sec:app_vis}
The visual branch represents the ECAL surface in a two-dimensional image domain by mapping the spherical coordinates $(\phi,\,\theta)$ to image coordinates $(x,\,y)$. The image resolution is set to $960\times480$ pixels, chosen so that ECAL cells with different angular sizes can be rasterized onto integer pixel regions as much as possible. Each calorimeter cell is therefore represented by a rectangular pixel region according to its angular coverage, preserving a piecewise uniform mapping from the detector surface to the image plane.

Because the deposited-energy distribution is highly imbalanced, the cell energy is first transformed by a base-10 logarithm and then encoded into three color channels using fixed dynamic-range intervals. Low-, medium- and high-energy deposits are assigned to the blue, green and red channels, respectively, with nonlinear rescaling within each interval to enhance weak deposits while avoiding saturation of high-energy cells. During training, a weak Gaussian black-background component is added as an image-level augmentation to improve robustness, whereas a deterministic seeded background is used for evaluation.

To encode the statistically averaged spatial structure of antineutron responses, the visual branch incorporates a physics-inspired diffusion operator~\cite{vheat2025} whose mathematical form follows a two-dimensional heat conduction equation,
\begin{equation}
    \frac{\partial u}{\partial t}
    = k \left( \frac{\partial^2 u}{\partial x^2}
    + \frac{\partial^2 u}{\partial y^2} \right),
    \label{eq:heat}
\end{equation}
where $k$ denotes the diffusion coefficient and $t$ the diffusion time. Given the initial condition $f(x,y)=u(x,y,t)|_{t=0}$, applying the Fourier transform gives the frequency-domain solution
\begin{equation}
    \tilde{u}(\omega_x,\omega_y,t)
    = \tilde{f}(\omega_x,\omega_y)
      e^{-k(\omega_x^2+\omega_y^2)t},
    \label{eq:heat_solve1}
\end{equation}
and hence
\begin{equation}
    u(x,y,t)
    = \mathcal{F}^{-1}\left[
      \tilde{f}(\omega_x,\omega_y)
      e^{-k(\omega_x^2+\omega_y^2)t}
      \right].
    \label{eq:heat_solve2}
\end{equation}
When applied to feature maps, $\mathcal{F}$ and $\mathcal{F}^{-1}$ are implemented using the two-dimensional discrete cosine transform (DCT) and inverse DCT (IDCT), yielding
\begin{equation}
    U^t =
    \operatorname{IDCT}_{2D}\left[
    \operatorname{DCT}_{2D}(U^0)
    e^{-k(\omega_x^2+\omega_y^2)t}
    \right],
    \label{eq:hco}
\end{equation}
where $U^0$ and $U^t$ denote the input and output feature maps, respectively. This operation provides an isotropic diffusion-like propagation of local activations and embeds the averaged centrality-and-diffusion prior directly into feature extraction.

The diffusion coefficient is parameterized rather than fixed, allowing the visual backbone to capture both shared diffusion characteristics and event-dependent variations. The resulting diffusion blocks are stacked hierarchically following a Swin-like design~\cite{swin}. The backbone combines Transformer-style components~\cite{attention}, including linear projections, layer normalization and feed-forward networks, with depth-wise convolution and a SiLU branch.

For detection-style learning, the visual branch uses RetinaNet~\cite{retinanet} as the base detector and constructs a multi-scale feature pyramid ($P_3$--$P_7$) through an FPN~\cite{fpn} built on backbone features at multiple resolutions ($C_3$--$C_5$). Because the reconstruction target is an incident point rather than a physical object extent, pseudo ground-truth bounding boxes are introduced for supervision. These boxes are centered on the labeled incident position, and their sizes are defined as multiples of the effective ECAL cell size. A two-dimensional Gaussian fit to the averaged particle image motivates a representative box size of 10 times of the effective cell size, approximately corresponding to a $\pm3\sigma$ interval.

The visual head predicts particle identity and incident direction through the detection pathway. Momentum magnitude is predicted by an auxiliary momentum head that aggregates broader contextual features; in multi-object settings, local FPN-level momentum predictions can also be used to complement the global estimate. The same formulation supports multi-object neutral-particle reconstruction and, when required, an auxiliary event-level classification output.

\subsection{Sequential representation branch}\label{sec:app_seq}
The sequential branch encodes each event as a fixed-length token sequence constructed from the set of deposited energy points on the ECAL surface. Each deposited point is represented by its energy and angular coordinates $(E,\,\phi,\,\theta)$, with the energy embedded on a logarithmic scale. As the number of deposited points $N$ varies event by event, the sequence is aligned to a fixed length $L=640$ by repeating the hit set and padding the remaining positions with all-zero tokens. Specifically, the repetition count is defined as $m=\lfloor L/N \rfloor$, producing $mN$ valid hit tokens. Within each repeated segment, the token order is randomly shuffled during training to avoid imposing an artificial ordering on the unordered hit set, and the azimuthal coordinate is cyclically shifted by $2\pi i/m$ for the $i$-th repetition, with wrapping back into $(-\pi,\,\pi)$ when necessary.

The input variables are discretized to fine bins and mapped to learnable embeddings. The energy is divided to 768 logarithmic bins and embedded into 768 channels, while $\phi$ and $\theta$ are divided into 360 and 180 angular bins and embedded into 384 channels each. The $\phi$--$\theta$ embeddings are concatenated to form an angular-position feature and added to the energy embedding. A learnable momentum token is appended to the sequence to aggregate event-level information for momentum regression and, when required, global event classification. Self-attention is controlled by an explicit mask: valid hit tokens interact within the same repeated segment, padded positions are isolated from valid hits, and the momentum token exchanges information only with valid deposits rather than padding artifacts.

Incident-point inference is formulated as an internal voting problem. Each deposited point acts as an anchor at $(\phi_{\rm anchor},\,\theta_{\rm anchor})$ and predicts both a confidence score and an angular offset $(\phi_{\rm out},\,\theta_{\rm out})$, decoded as
\begin{equation}
\phi_{\rm pred}=\phi_{\rm anchor}+\phi_{\rm base}\phi_{\rm out}, \qquad
\theta_{\rm pred}=\theta_{\rm anchor}+\theta_{\rm base}\theta_{\rm out},
\end{equation}
where $\phi_{\rm base}$ and $\theta_{\rm base}$ define the angular regression scales. During training, anchors close to the labeled incident point are assigned as positive examples. The confidence branch is supervised by a focal loss, and the angular offsets are supervised by L1 losses on the encoded targets. During inference, each valid token produces a candidate incident point, and the highest-confidence proposal is selected as the event-level direction estimate.

Momentum magnitude is regressed once per event from the momentum token using a direct residual parameterization,
\begin{equation}
p_{\rm pred}=p_{\rm out}\sigma_p+\mu_p+p_{\rm base},
\end{equation}
where $p_{\rm base}$ is a dataset-scale momentum prior and $\mu_p,\sigma_p$ denote the normalization constants. In the configurations used in this study, $\mu_p=0$ and $\sigma_p=1$, so the network directly predicts a residual relative to $p_{\rm base}$. When the global classification task is enabled, an auxiliary event-level classification output is trained jointly with the direction and momentum objectives.

To improve representation capacity without requiring incident labels, the sequential branch employs a position-cloze pretraining objective tailored to EMC hit sequences rather than natural-language tokens or regular image patches~\cite{devlin2019bert,mae}. The task masks the angular coordinates of selected informative hit tokens and asks a decoder to recover $(\phi,\,\theta)$ from the unmasked hit context and retained energy prompts. Recovering positions, rather than energy values, focuses on relatively stable geometric patterns, as hit energies fluctuate strongly under hadronic interaction. In practice, masked tokens are sampled from the top 50\% of deposited energies with a masking ratio of 20\%. The pretraining network uses a 12-layer encoder with 768-dimensional features and an 8-layer decoder with 384-dimensional features, and the loss is applied only to the masked positions. For downstream fine-tuning, the network reuses both the pretrained encoder and decoder. 

\subsection{Mixed representation framework}\label{sec:app_mix}
The mixed-representation model receives two parallel inputs constructed from the same ECAL event: the rasterized image used by the visual branch and the fixed-length hit-token sequence used by the sequential branch. The two branches are encoded in parallel, preserving their distinct inductive biases. The visual branch provides multi-scale image features for detection-style reconstruction, whereas the sequential branch provides contextualized hit-token features and a momentum token for event-level aggregation. Their outputs are passed to a unified prediction interface that uses common conventions for particle class, incident direction, momentum magnitude and optional global classification.

During training, both branches are supervised with the same reconstruction targets. The mixed model is optimized by combining branch-specific detection losses, angular-regression losses and momentum-regression losses. When the global classification task is enabled, the corresponding cross-entropy loss is added to the joint objective. This training strategy allows each representation branch to learn from the same physical labels while retaining its own feature-extraction pathway.

During inference, the visual and sequential branches first generate candidate detections independently. Each candidate contains a particle label, a confidence score, an incident direction represented by $(\phi,\theta)$, and a momentum-magnitude estimate. Because the two branches may have different score calibrations, candidate scores are first calibrated with branch-dependent temperature factors,
\begin{equation}
    \bar{s}=\sigma\left(\frac{\operatorname{logit}(s)}{T_{b,c}}\right),
\end{equation}
where $s$ is the raw confidence score, $T_{b,c}$ is the temperature for branch $b$ and class $c$, and $\sigma$ denotes the sigmoid function.

Candidates from the two branches are then matched within each particle class according to their angular agreement and calibrated confidence. For a visual candidate $i$ and a sequential candidate $j$, the angular distance is computed on the unit sphere as
\begin{equation}
\Delta\Omega_{ij} =
\arccos\left[
\sin\theta_i\sin\theta_j\cos(\phi_i-\phi_j)
+\cos\theta_i\cos\theta_j
\right].
\end{equation}
Candidate pairs outside a class-dependent angular matching threshold are excluded. For the remaining pairs, a matching cost combines angular distance and confidence,
\begin{equation}
    C_{ij}=\Delta\Omega_{ij}
    -\lambda\left(\bar{s}_i^{\rm vis}+\bar{s}_j^{\rm seq}\right),
\end{equation}
where $\lambda$ controls the relative contribution of confidence. A global one-to-one assignment is then obtained for each class.

Matched candidates within a tighter angular fusion threshold are merged into a single prediction. The incident directions are converted to unit vectors and averaged with confidence weights,
\begin{equation}
    \hat{\mathbf{n}} =
    \frac{
    w_{\rm vis}\mathbf{n}_{\rm vis}
    +w_{\rm seq}\mathbf{n}_{\rm seq}}
    {\left\|
    w_{\rm vis}\mathbf{n}_{\rm vis}
    +w_{\rm seq}\mathbf{n}_{\rm seq}
    \right\|},
    \qquad
    w_b=\frac{\bar{s}_b}{\bar{s}_{\rm vis}+\bar{s}_{\rm seq}},
\end{equation}
and the momentum magnitude is fused using the same weights. The fused confidence score is computed as the probability union of the two branch scores. Matched pairs that do not satisfy the fusion threshold are resolved by retaining the higher-confidence branch prediction, and unmatched high-confidence candidates are preserved. The resulting candidate set is finally used as the unified MrCAL output for downstream reconstruction.

\section{Supplementary figures}\label{sec:suppfig}
\begin{figure}[h]
    \centering
    \footnotesize
    \begin{minipage}[b]{0.55\linewidth}
        \centering
        \includegraphics[width=\linewidth]{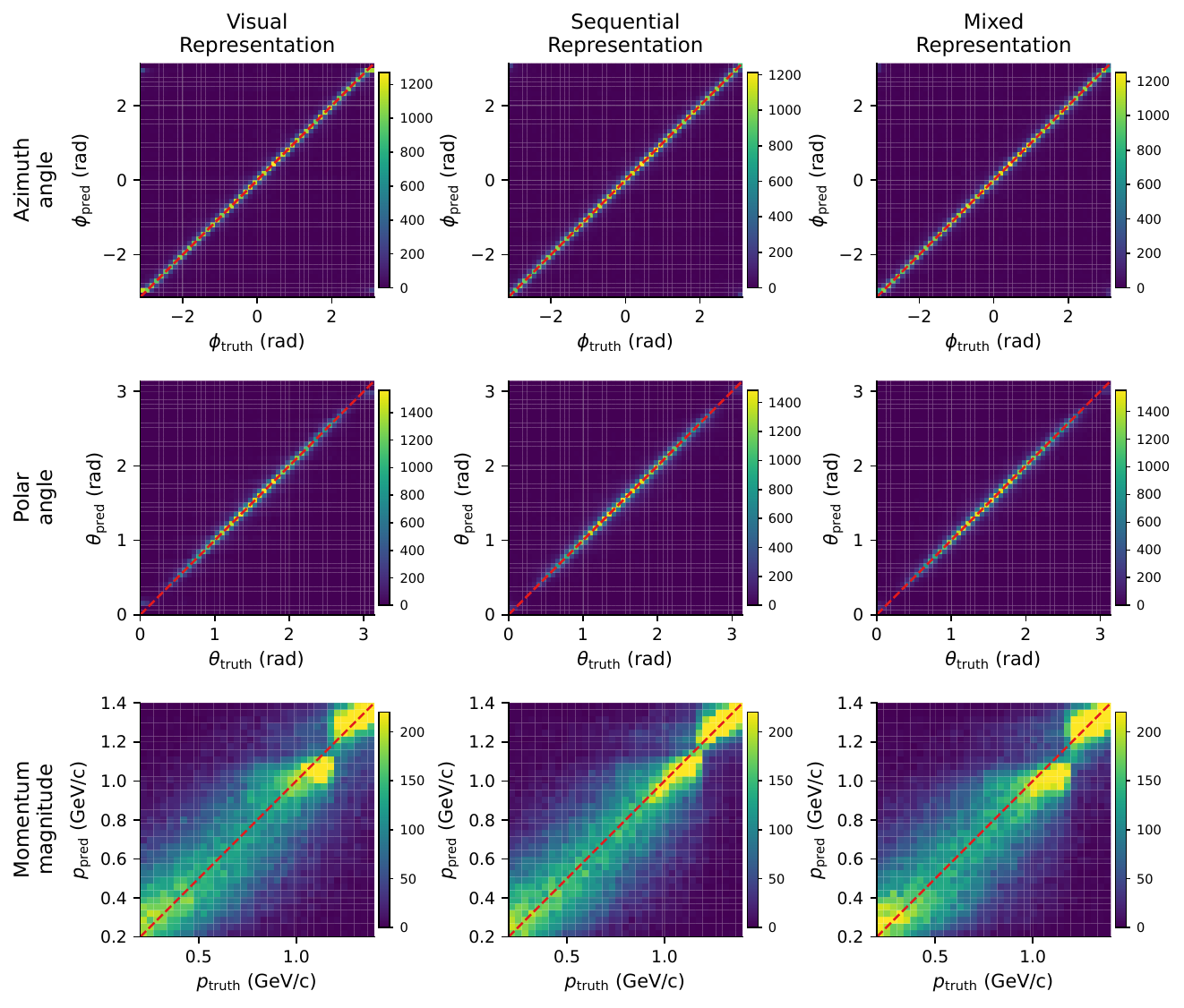}
        \textbf{(a)}
    \end{minipage}
    \begin{minipage}[b]{0.44\linewidth}
        \centering
        \includegraphics[width=\linewidth]{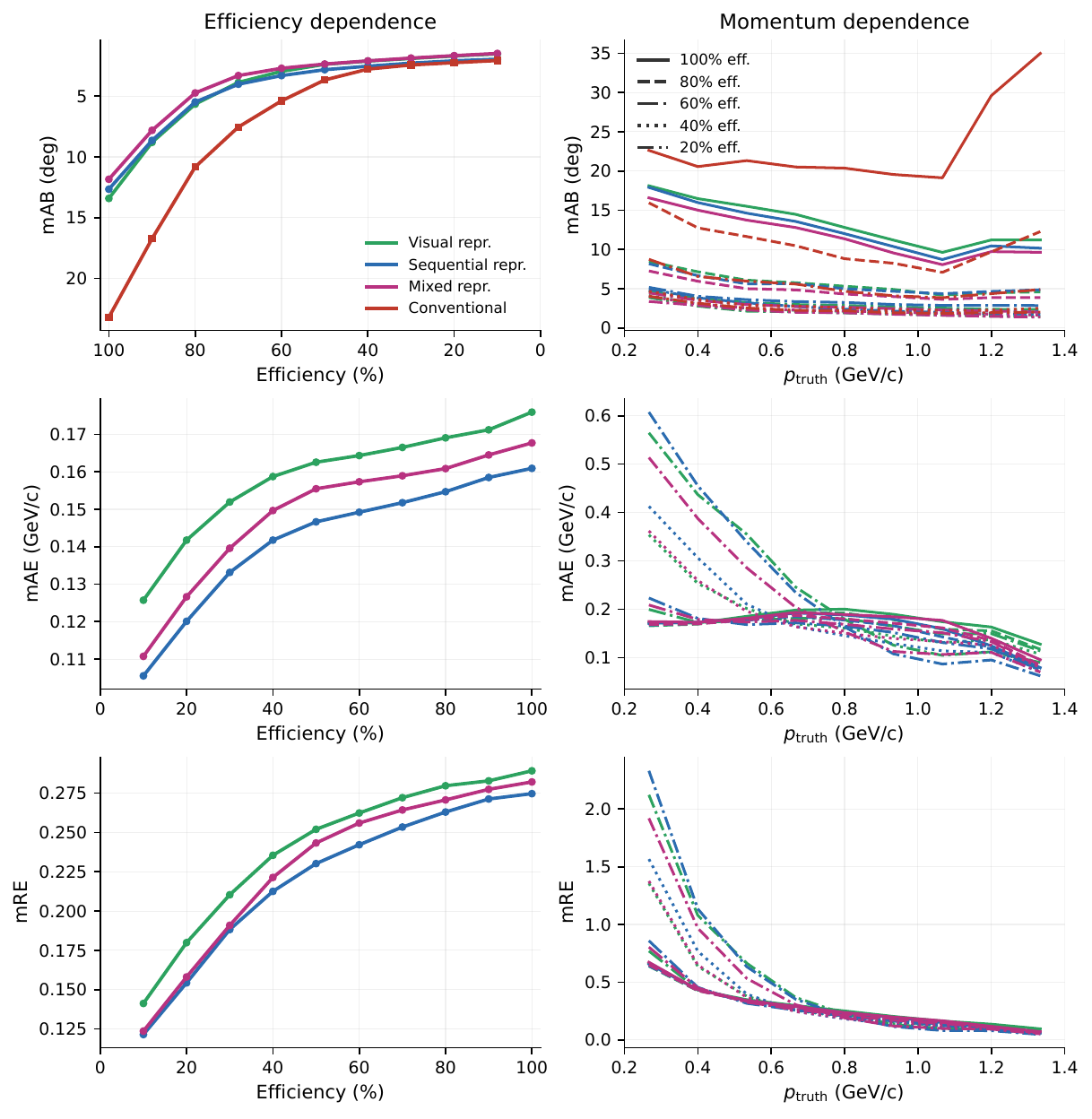}
        \textbf{(b)}
    \end{minipage}
    \caption{Performance evaluation in the single-particle setting. (a) Two-dimensional correlations between the true and predicted azimuthal angle, polar angle, and momentum magnitude of the antineutron. (b) Dependence of the mAB, mAE, and mRE metrics on efficiency and on the true momentum magnitude.}
    \label{fig:supp_eval_single}
\end{figure}

\begin{figure}[h]
    \centering
    \footnotesize
    \begin{minipage}[b]{0.55\linewidth}
        \centering
        \includegraphics[width=\linewidth]{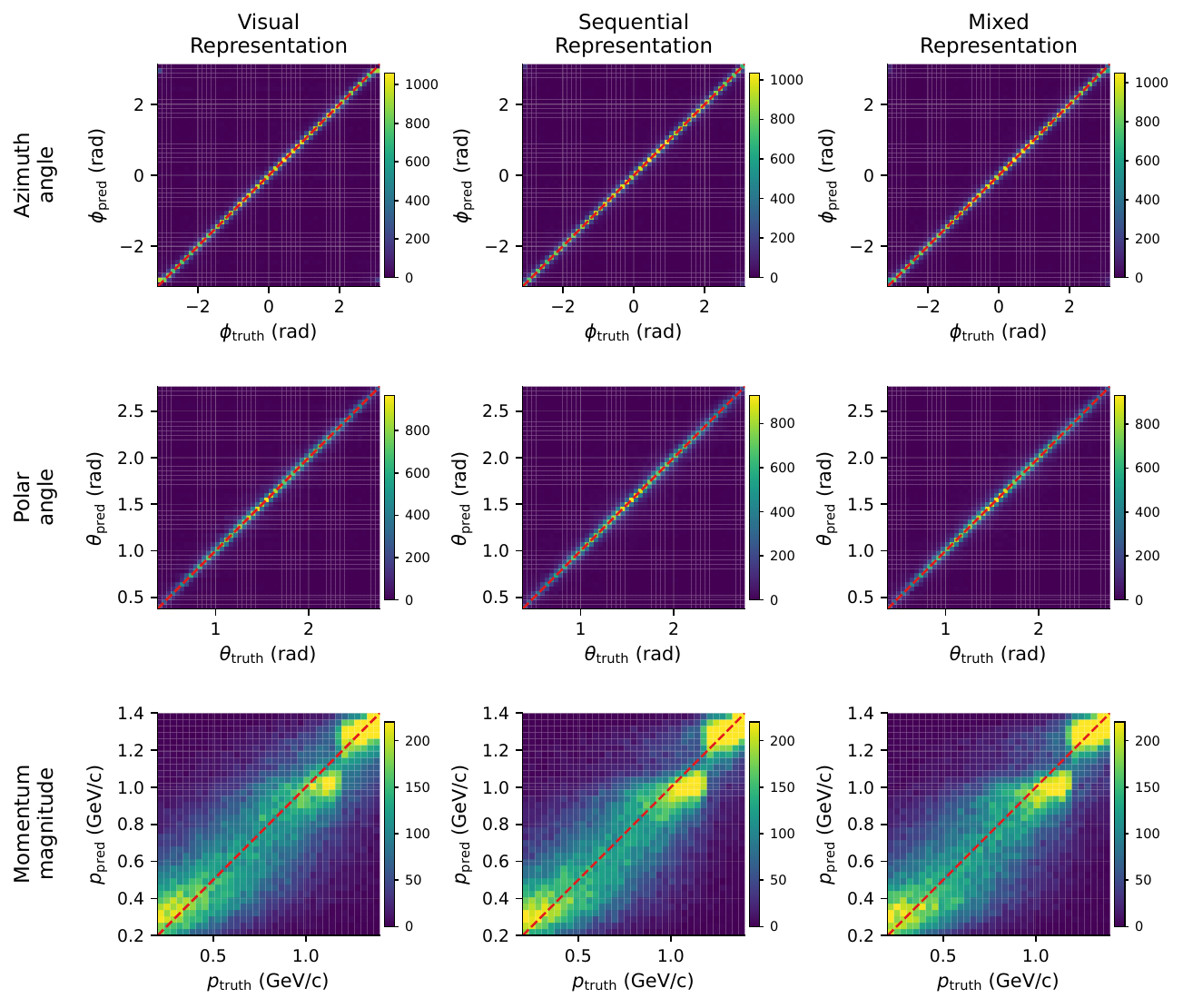}
        \textbf{(a)}
    \end{minipage}
    \begin{minipage}[b]{0.44\linewidth}
        \centering
        \includegraphics[width=\linewidth]{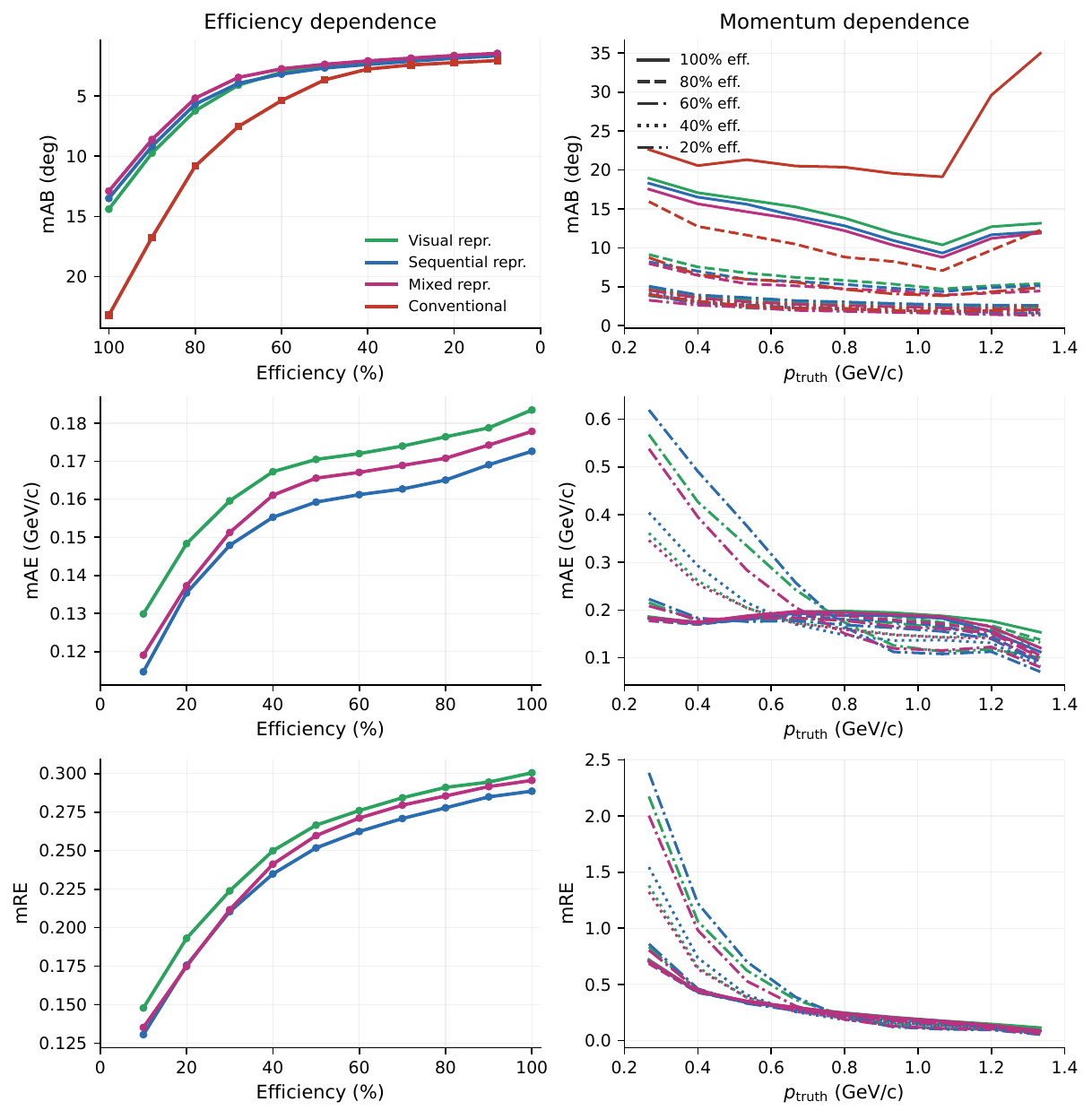}
        \textbf{(b)}
    \end{minipage}
    \caption{Performance evaluation in the multi-particle setting for the antineutron. (a) Two-dimensional correlations between the true and predicted azimuthal angle, polar angle, and momentum magnitude of the antineutron. (b) Dependence of the mAB, mAE, and mRE metrics on efficiency and on the true momentum magnitude.}
    \label{fig:supp_eval_multi_n}
\end{figure}

\begin{figure}[h]
    \centering
    \footnotesize
    \begin{minipage}[b]{0.55\linewidth}
        \centering
        \includegraphics[width=\linewidth]{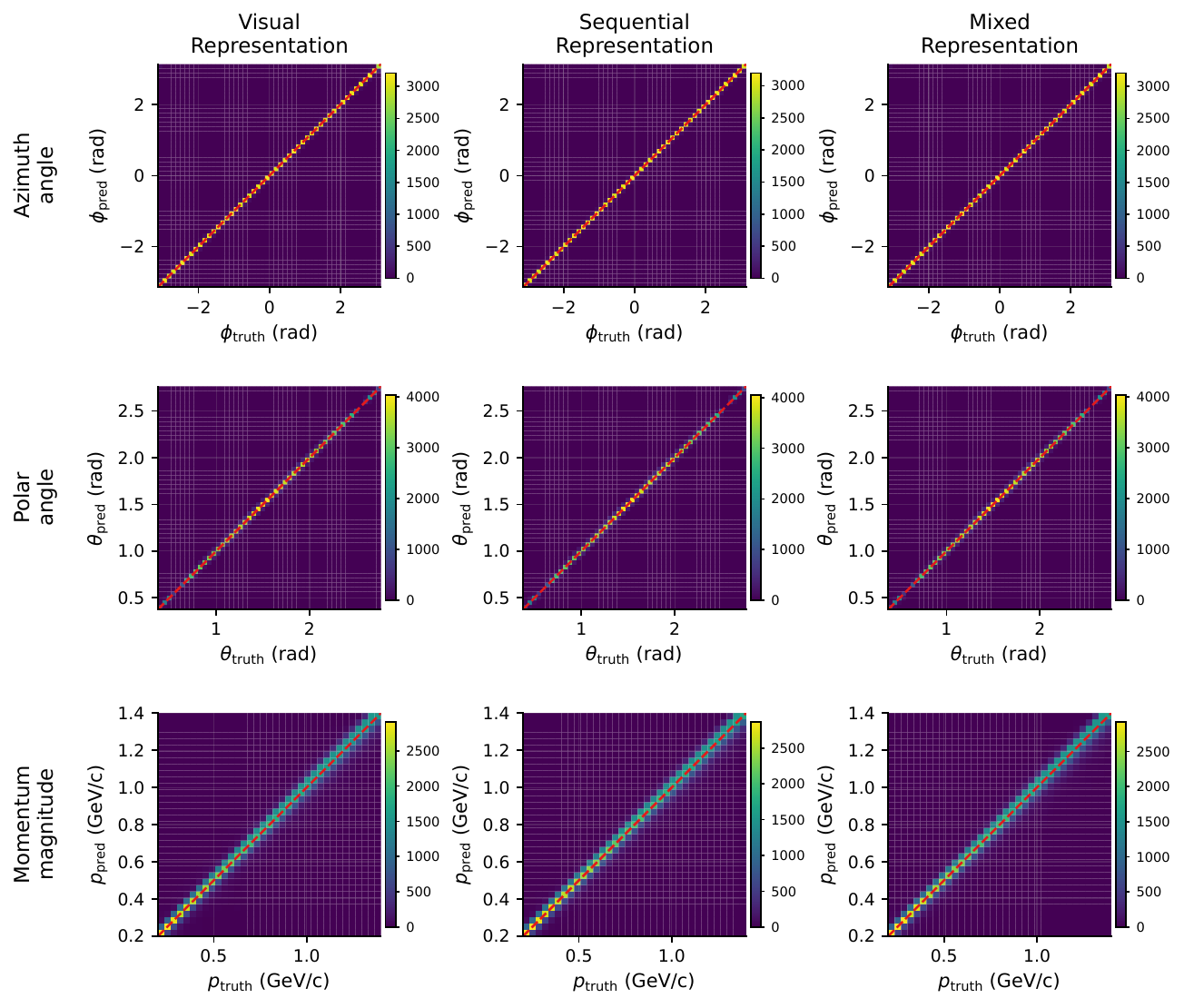}
        \textbf{(a)}
    \end{minipage}
    \begin{minipage}[b]{0.44\linewidth}
        \centering
        \includegraphics[width=\linewidth]{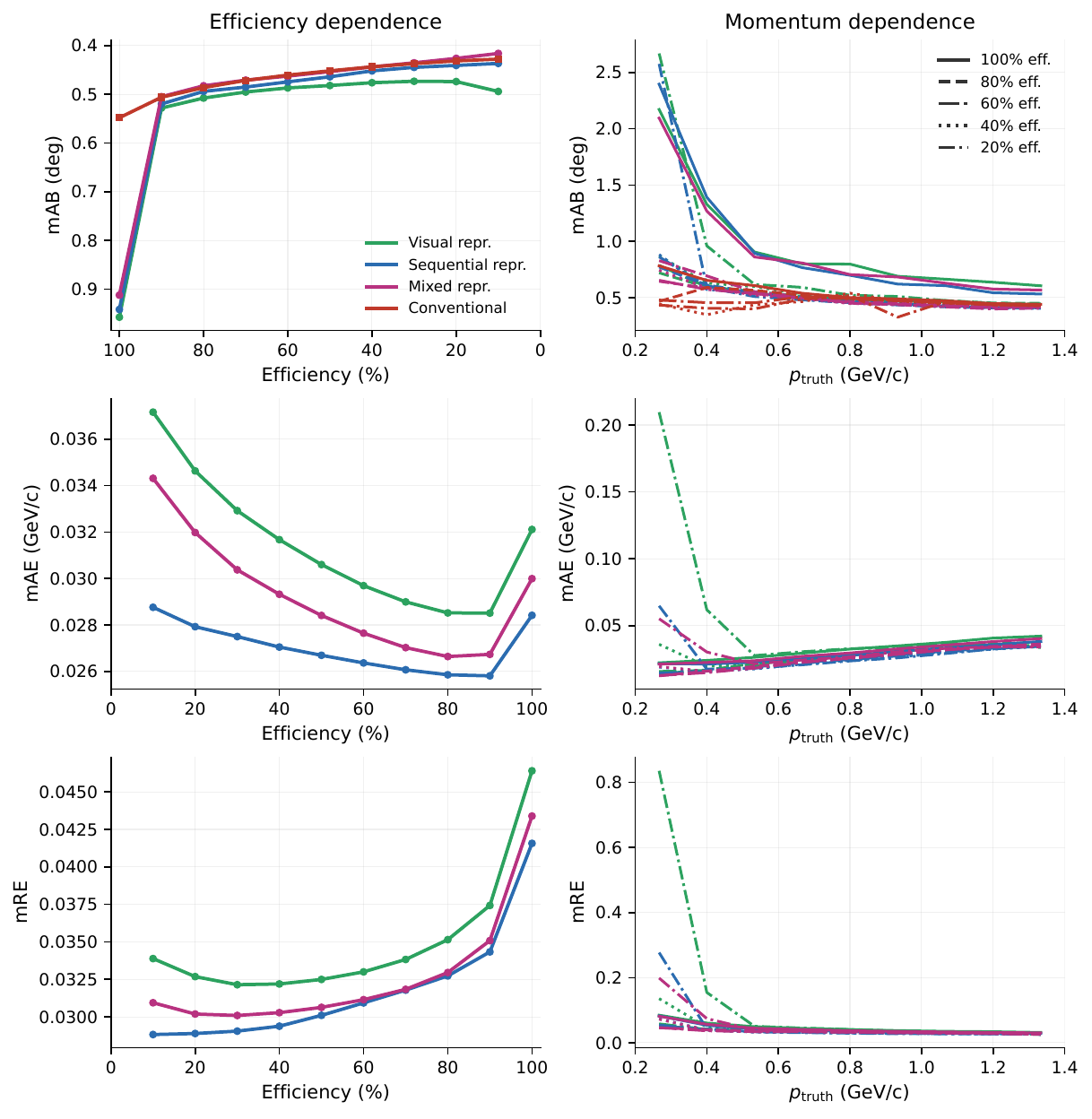}
        \textbf{(b)}
    \end{minipage}
    \caption{Performance evaluation in the multi-particle setting for the photon. (a) Two-dimensional correlations between the true and predicted azimuthal angle, polar angle, and momentum magnitude of the antineutron. (b) Dependence of the mAB, mAE, and mRE metrics on efficiency and on the true momentum magnitude.}
    \label{fig:supp_eval_multi_g}
\end{figure}

\begin{figure}[h]
    \centering
    \includegraphics[width=\linewidth]{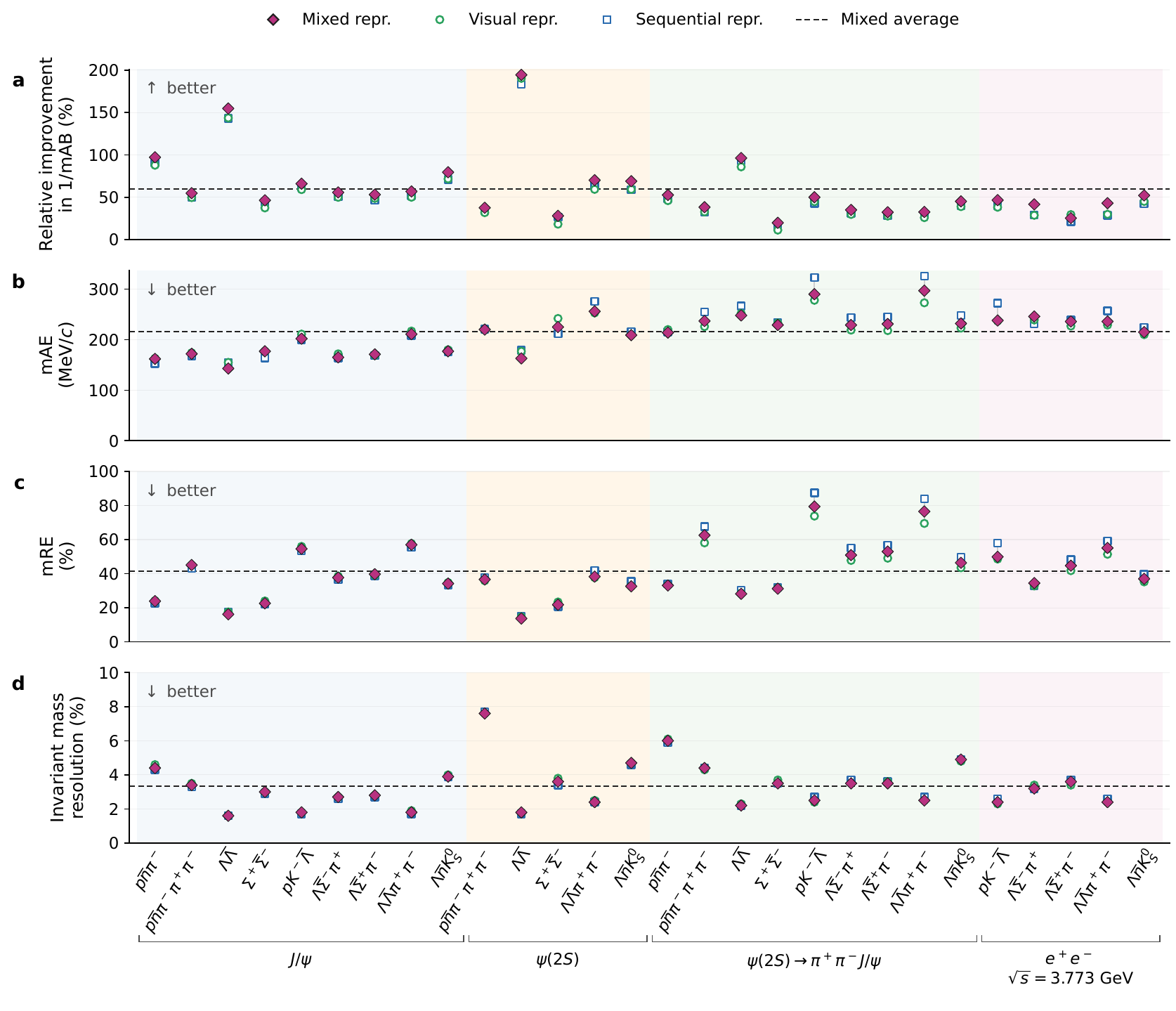}
    \caption{Generalization test in the single-particle setting. Comparisons of the relative improvement in (a) directional precision, (b) the mAE, (c) the mRE and (d) the invariant mass resolution across benchmarking channels.}
    \label{fig:supp_gene_single}
\end{figure}

\begin{figure}[h]
    \centering
    \includegraphics[width=0.8\linewidth]{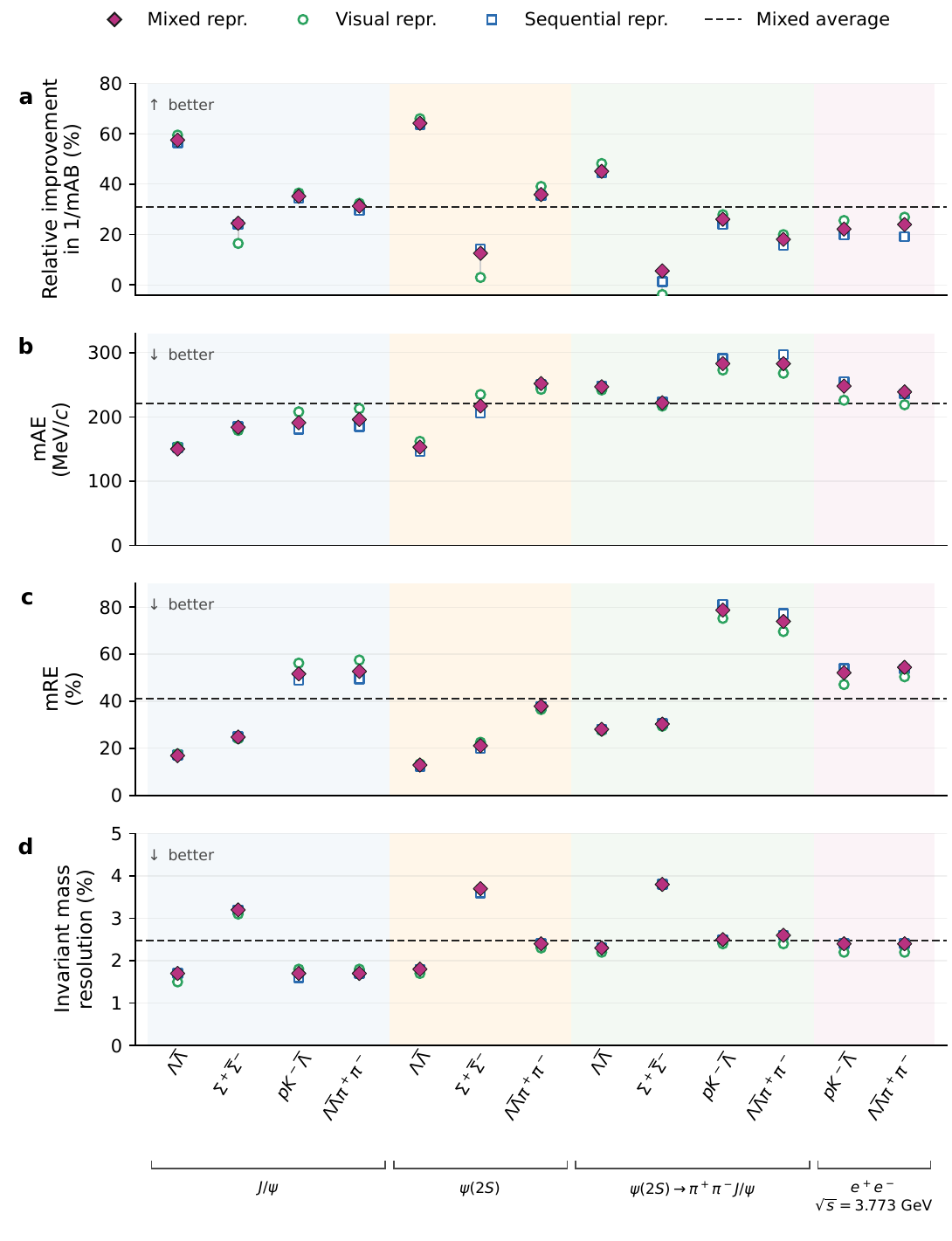}
    \caption{Generalization test in the multi-particle setting for the antineutron. Comparisons of the relative improvement in (a) directional precision, (b) the mAE, (c) the mRE and (d) the invariant mass resolution across benchmarking channels.}
    \label{fig:supp_gene_multi_n}
\end{figure}

\begin{figure}[h]
    \centering
    \includegraphics[width=0.8\linewidth]{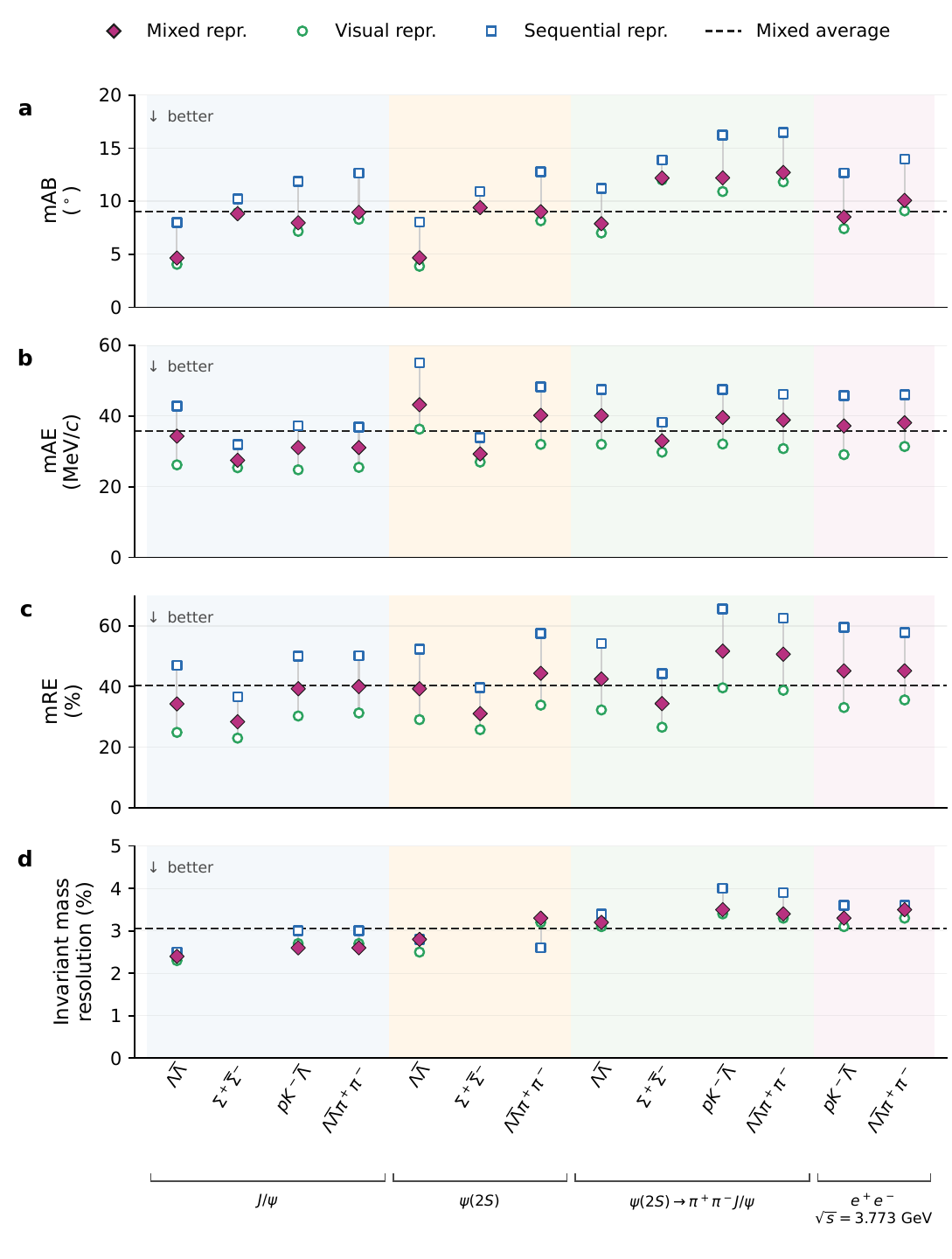}
    \caption{Generalization test in the multi-particle setting for the photon. Comparisons of the relative improvement in (a) directional precision, (b) the mAE, (c) the mRE and (d) the invariant mass resolution across benchmarking channels.}
    \label{fig:supp_gene_multi_g}
\end{figure}

\begin{figure}[h]
    \centering
    \includegraphics[width=\linewidth]{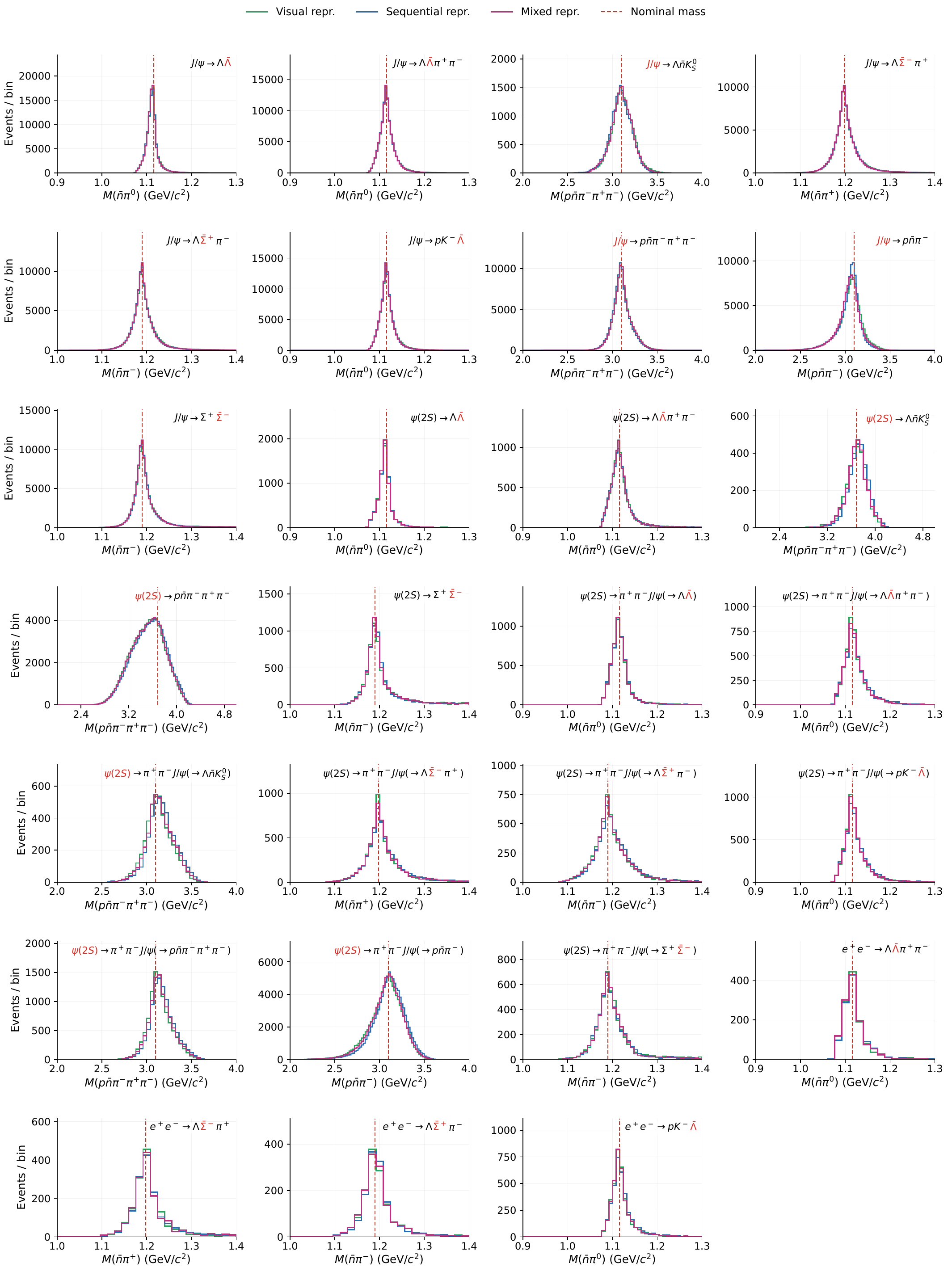}
    \caption{Generalization test in the single-particle setting. Invariant mass spectra of the reconstructed short-lived particles decaying to antineutron. Types and positions of these particles in their decay chains are highlighted in red.}
    \label{fig:supp_gene_invm_single}
\end{figure}

\begin{figure}[h]
    \centering
    \includegraphics[width=\linewidth]{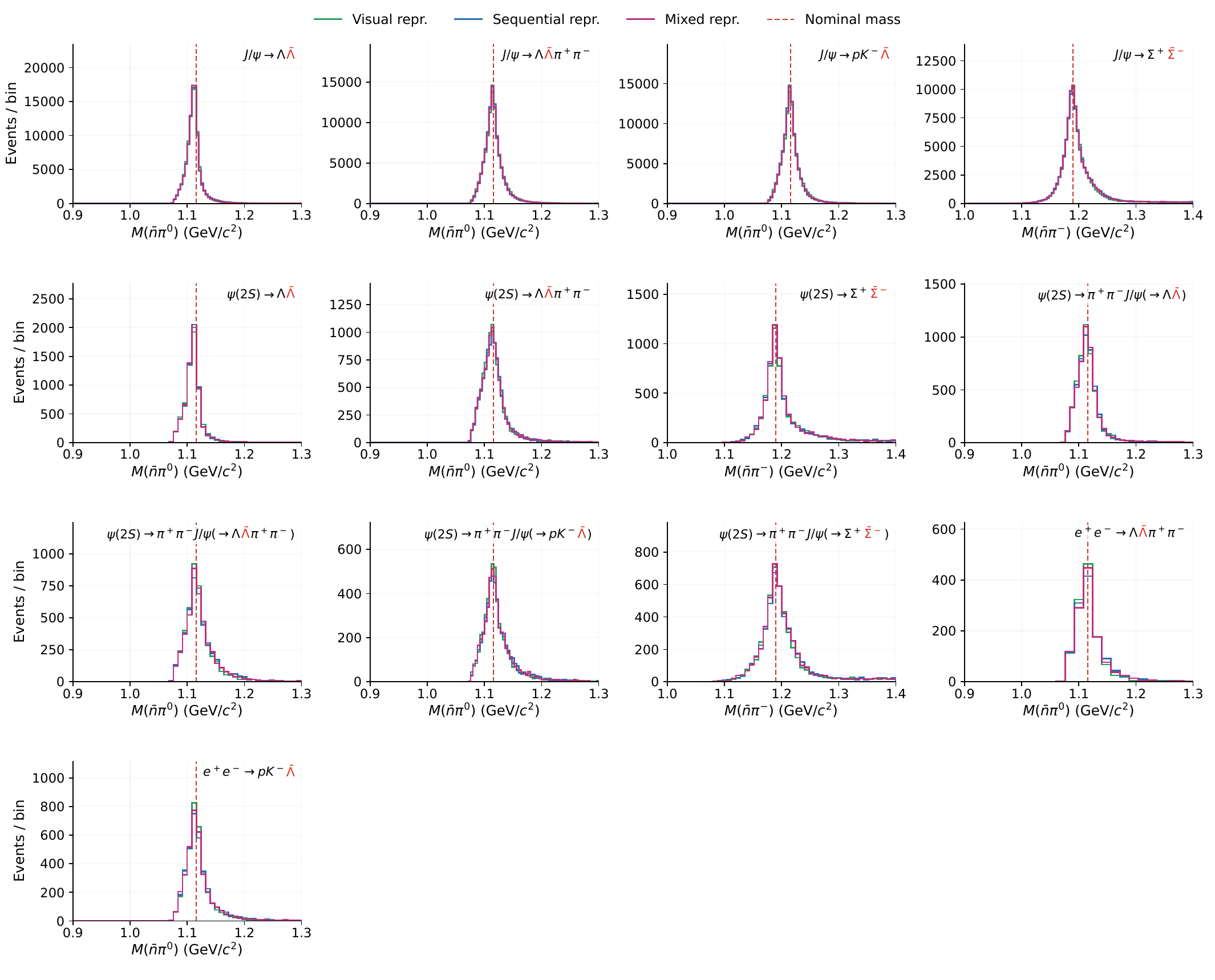}
    \caption{Generalization test in the multi-particle setting. Invariant mass spectra of the reconstructed short-lived particles decaying to antineutron. Types and positions of these particles in their decay chains are highlighted in red.}
    \label{fig:supp_gene_invm_multi}
\end{figure}

\clearpage
\bibliographystyle{JHEP}
\bibliography{bibitem}

@article{retinanet,
  title={Focal Loss for Dense Object Detection},
  author={Lin, Tsung-Yi and Goyal, Priya and Girshick, Ross and He, Kaiming and Doll{\'a}r, Piotr},
  journal={IEEE Transactions on Pattern Analysis and Machine Intelligence},
  volume={42},
  number={2},
  pages={318--327},
  year={2018},
  publisher={IEEE},
  doi={10.1109/TPAMI.2018.2858826},
}

@inproceedings{fpn,
  title={Feature pyramid networks for object detection},
  author={Lin, Tsung-Yi and Doll{\'a}r, Piotr and Girshick, Ross and He, Kaiming and Hariharan, Bharath and Belongie, Serge},
  booktitle={Proceedings of the IEEE conference on computer vision and pattern recognition},
  pages={2117--2125},
  year={2017},
  doi={10.1109/CVPR.2017.106},
}

@article{attention,
  title={Attention is all you need},
  author={Vaswani, Ashish and Shazeer, Noam and Parmar, Niki and Uszkoreit, Jakob and Jones, Llion and Gomez, Aidan N and Kaiser, {\L}ukasz and Polosukhin, Illia},
  journal={Advances in neural information processing systems},
  volume={30},
  year={2017},
  doi={10.48550/arXiv.1706.03762}
}

@inproceedings{swin,
  title={Swin transformer: Hierarchical vision transformer using shifted windows},
  author={Liu, Ze and Lin, Yutong and Cao, Yue and Hu, Han and Wei, Yixuan and Zhang, Zheng and Lin, Stephen and Guo, Baining},
  booktitle={Proceedings of the IEEE/CVF international conference on computer vision},
  pages={10012--10022},
  year={2021},
  doi={10.1109/ICCV48922.2021.00986},
}

@inproceedings{vheat2025,
  title={Building Vision Models upon Heat Conduction},
  author={Wang, Zhaozhi and Liu, Yue and Tian, Yunjie and Liu, Yunfan and Wang, Yaowei and Ye, Qixiang},
  booktitle={Proceedings of the Computer Vision and Pattern Recognition Conference},
  pages={9707--9717},
  year={2025},
  doi={10.1109/CVPR52734.2025.00907},
}

@article{voc,
  title={The pascal visual object classes (voc) challenge},
  author={Everingham, Mark and Van Gool, Luc and Williams, Christopher KI and Winn, John and Zisserman, Andrew},
  journal={International journal of computer vision},
  volume={88},
  number={2},
  pages={303--338},
  year={2010},
  publisher={Springer},
  doi={10.1007/S11263-009-0275-4},
}

@inproceedings{mscoco,
  title={Microsoft coco: Common objects in context},
  author={Lin, Tsung-Yi and Maire, Michael and Belongie, Serge and Hays, James and Perona, Pietro and Ramanan, Deva and Doll{\'a}r, Piotr and Zitnick, C Lawrence},
  booktitle={European conference on computer vision},
  pages={740--755},
  year={2014},
  organization={Springer},
  doi={10.1007/978-3-319-10602-1\_48},
}

@inproceedings{mae,
  title={Masked autoencoders are scalable vision learners},
  author={He, Kaiming and Chen, Xinlei and Xie, Saining and Li, Yanghao and Doll{\'a}r, Piotr and Girshick, Ross},
  booktitle={Proceedings of the IEEE/CVF conference on computer vision and pattern recognition},
  pages={16000--16009},
  year={2022},
  doi={10.1109/CVPR52688.2022.01553},
}

@inproceedings{devlin2019bert,
  title={Bert: Pre-training of deep bidirectional transformers for language understanding},
  author={Devlin, Jacob and Chang, Ming-Wei and Lee, Kenton and Toutanova, Kristina},
  booktitle={Proceedings of the 2019 conference of the North American chapter of the association for computational linguistics: human language technologies, volume 1 (long and short papers)},
  pages={4171--4186},
  year={2019},
  doi={10.18653/V1/N19-1423},
}

@article{YUSSUP2026113377,
title = {A decade of trends and progress in methods and hardware accelerators for particle track reconstruction: a systematic review},
journal = {Radiation Physics and Chemistry},
volume = {239},
pages = {113377},
year = {2026},
issn = {0969-806X},
doi = {10.1016/j.radphyschem.2025.113377},
url = {https://www.sciencedirect.com/science/article/pii/S0969806X25008692},
author = {Nolida Yussup and Mohd Idzat Idris and Imran Yusuff}
}

@book{Fruhwirth:2020zbo,
    author = {Fr{\"u}hwirth, Rudolf and Strandlie, Are},
    title = "Pattern Recognition, Tracking and Vertex Reconstruction in Particle Detectors",
    doi = "10.1007/978-3-030-65771-0",
    isbn = "978-3-030-65770-3, 978-3-030-65771-0",
    publisher = "Springer",
    series = "Particle Acceleration and Detection",
    month = "11",
    year = "2020"
}

@article{BESIII:2022rgl,
    author = "Ablikim, M. and others",
    collaboration = "BESIII",
    title = "{Measurement of the Absolute Branching Fraction and Decay Asymmetry of {\ensuremath{\Lambda}}{\textrightarrow}n{\ensuremath{\gamma}}}",
    eprint = "2206.10791",
    archivePrefix = "arXiv",
    primaryClass = "hep-ex",
    doi = "10.1103/PhysRevLett.129.212002",
    journal = "Phys. Rev. Lett.",
    volume = "129",
    number = "21",
    pages = "212002",
    year = "2022"
}

@article{BESIII:2015jmz,
    author = "Ablikim, M. and others",
    collaboration = "BESIII",
    title = "{Study of decay dynamics and $CP$ asymmetry in $D^+ \to K^0_L e^+ \nu_e$ decay}",
    eprint = "1510.00308",
    archivePrefix = "arXiv",
    primaryClass = "hep-ex",
    doi = "10.1103/PhysRevD.92.112008",
    journal = "Phys. Rev. D",
    volume = "92",
    number = "11",
    pages = "112008",
    year = "2015"
}

@article{BESIII:2021tbq,
    author = "Ablikim, Medina and others",
    collaboration = "BESIII",
    title = "{Oscillating features in the electromagnetic structure of the neutron}",
    eprint = "2103.12486",
    archivePrefix = "arXiv",
    primaryClass = "hep-ex",
    doi = "10.1038/s41567-021-01345-6",
    journal = "Nature Phys.",
    volume = "17",
    number = "11",
    pages = "1200--1204",
    year = "2021"
}

@article{ParticleDataGroup:2024cfk,
    author = "Navas, S. and others",
    collaboration = "Particle Data Group",
    title = "{Review of particle physics}",
    doi = "10.1103/PhysRevD.110.030001",
    journal = "Phys. Rev. D",
    volume = "110",
    number = "3",
    pages = "030001",
    year = "2024"
}

@article{BaBar:2001yhh,
    author = "Aubert, Bernard and others",
    collaboration = "BaBar",
    title = "{The BaBar detector}",
    eprint = "hep-ex/0105044",
    archivePrefix = "arXiv",
    reportNumber = "SLAC-PUB-8569, BABAR-PUB-01-08",
    doi = "10.1016/S0168-9002(01)02012-5",
    journal = "Nucl. Instrum. Meth. A",
    volume = "479",
    pages = "1--116",
    year = "2002"
}

@article{CLEO:1991qyy,
    author = "Kubota, Y. and others",
    collaboration = "CLEO",
    title = "{The CLEO-II detector}",
    reportNumber = "CLNS-91-1122, CLEO-91-11",
    doi = "10.1016/0168-9002(92)90770-5",
    journal = "Nucl. Instrum. Meth. A",
    volume = "320",
    pages = "66--113",
    year = "1992"
}

@article{Belle-II:2018jsg,
    author = "Altmannshofer, W. and others",
    editor = "Kou, E. and Urquijo, P.",
    collaboration = "Belle-II",
    title = "{The Belle II Physics Book}",
    eprint = "1808.10567",
    archivePrefix = "arXiv",
    primaryClass = "hep-ex",
    reportNumber = "KEK Preprint 2018-27, BELLE2-PUB-PH-2018-001, FERMILAB-PUB-18-398-T, JLAB-THY-18-2780, INT-PUB-18-047, UWThPh 2018-26",
    doi = "10.1093/ptep/ptz106",
    journal = "PTEP",
    volume = "2019",
    number = "12",
    pages = "123C01",
    year = "2019",
    note = "[Erratum: PTEP 2020, 029201 (2020)]"
}

@article{Achasov:2023gey,
    author = "Achasov, M. and others",
    title = "{STCF conceptual design report (Volume 1): Physics {\&} detector}",
    eprint = "2303.15790",
    archivePrefix = "arXiv",
    primaryClass = "hep-ex",
    doi = "10.1007/s11467-023-1333-z",
    journal = "Front. Phys. (Beijing)",
    volume = "19",
    number = "1",
    pages = "14701",
    year = "2024"
}

@article{Pietropaolo:2020frm,
    author = "Pietropaolo, A. and others",
    title = "{Neutron detection techniques from \ensuremath{\mu}eV to GeV}",
    doi = "10.1016/j.physrep.2020.06.003",
    journal = "Phys. Rept.",
    volume = "875",
    pages = "1--65",
    year = "2020"
}

@article{LIU2022166672,
    title = {Development of a data-driven method to simulate the detector response of anti-neutron at BESIII},
    journal = {Nuclear Instruments and Methods in Physics Research Section A: Accelerators, Spectrometers, Detectors and Associated Equipment},
    volume = {1033},
    pages = {166672},
    year = {2022},
    issn = {0168-9002},
    doi = {10.1016/j.nima.2022.166672},
    url = {https://www.sciencedirect.com/science/article/pii/S0168900222002194},
    author = {Liang Liu and Xiaorong Zhou and Haiping Peng}
}

@article{He:2011zzd,
    author = "He, Miao",
    editor = "Wang, Yifang",
    title = "{Simulation and reconstruction of the BESIII EMC}",
    doi = "10.1088/1742-6596/293/1/012025",
    journal = "J. Phys. Conf. Ser.",
    volume = "293",
    pages = "012025",
    year = "2011"
}

@article{BESIII:2024mgg,
    author = "Ablikim, Medina and others",
    collaboration = "BESIII",
    title = "{Observation of a rare beta decay of the charmed baryon with a Graph Neural Network}",
    eprint = "2410.13515",
    archivePrefix = "arXiv",
    primaryClass = "hep-ex",
    doi = "10.1038/s41467-024-55042-y",
    journal = "Nature Commun.",
    volume = "16",
    number = "1",
    pages = "681",
    year = "2025"
}

@article{BESIII:2009fln,
    author = "Ablikim, M. and others",
    collaboration = "BESIII",
    title = "{Design and Construction of the BESIII Detector}",
    eprint = "0911.4960",
    archivePrefix = "arXiv",
    primaryClass = "physics.ins-det",
    doi = "10.1016/j.nima.2009.12.050",
    journal = "Nucl. Instrum. Meth. A",
    volume = "614",
    pages = "345--399",
    year = "2010"
}

@article{GEANT4:2002zbu,
    author = "Agostinelli, S. and others",
    collaboration = "GEANT4",
    title = "{GEANT4--a simulation toolkit}",
    reportNumber = "SLAC-PUB-9350, FERMILAB-PUB-03-339, CERN-IT-2002-003",
    doi = "10.1016/S0168-9002(03)01368-8",
    journal = "Nucl. Instrum. Meth. A",
    volume = "506",
    pages = "250--303",
    year = "2003"
}

@article{BESIII:2021cxx,
    author = "Ablikim, M. and others",
    collaboration = "BESIII",
    title = "{Number of $J/\psi$ events at BESIII}",
    eprint = "2111.07571",
    archivePrefix = "arXiv",
    primaryClass = "hep-ex",
    doi = "10.1088/1674-1137/ac5c2e",
    journal = "Chin. Phys. C",
    volume = "46",
    number = "7",
    pages = "074001",
    year = "2022"
}

@article{BESIII:2024lks,
    author = "Ablikim, Medina and others",
    collaboration = "BESIII",
    title = "{Determination of the number of {\ensuremath{\psi}}(3686) events taken at BESIII*}",
    eprint = "2403.06766",
    archivePrefix = "arXiv",
    primaryClass = "hep-ex",
    doi = "10.1088/1674-1137/ad595b",
    journal = "Chin. Phys. C",
    volume = "48",
    number = "9",
    pages = "093001",
    year = "2024"
}

@article{BESIII:2024lbn,
    author = "Ablikim, Medina and others",
    collaboration = "BESIII",
    title = "{Measurement of integrated luminosity of data collected at 3.773 GeV by BESIII from 2021 to 2024*}",
    eprint = "2406.05827",
    archivePrefix = "arXiv",
    primaryClass = "hep-ex",
    doi = "10.1088/1674-1137/ad70a0",
    journal = "Chin. Phys. C",
    volume = "48",
    number = "12",
    pages = "123001",
    year = "2024"
}

@article{Cavallari_2011,
    doi = {10.1088/1742-6596/293/1/012001},
    url = {https://dx.doi.org/10.1088/1742-6596/293/1/012001},
    year = {2011},
    month = {apr},
    publisher = {},
    volume = {293},
    number = {1},
    pages = {012001},
    author = {Francesca Cavallari},
    title = {Performance of calorimeters at the LHC},
    journal = {Journal of Physics: Conference Series}
}

@article{Krause:2024avx,
    author = "Amram, Oz and others",
    editor = "Krause, Claudius and Faucci Giannelli, Michele and Kasieczka, Gregor and Nachman, Benjamin and Salamani, Dalila and Shih, David and Zaborowska, Anna",
    title = "{CaloChallenge 2022: a community challenge for fast calorimeter simulation}",
    eprint = "2410.21611",
    archivePrefix = "arXiv",
    primaryClass = "physics.ins-det",
    reportNumber = "HEPHY-ML-24-05, FERMILAB-PUB-24-0728-CMS, TTK-24-43",
    doi = "10.1088/1361-6633/ae1304",
    journal = "Rept. Prog. Phys.",
    volume = "88",
    number = "11",
    pages = "116201",
    year = "2025"
}

@article{Belle-II:2023cal,
    author = "Wemmer, F. and others",
    collaboration = "Belle-II",
    title = "{Photon Reconstruction in the Belle~II Calorimeter Using Graph Neural Networks}",
    eprint = "2306.04179",
    archivePrefix = "arXiv",
    primaryClass = "hep-ex",
    doi = "10.1007/s41781-023-00105-w",
    journal = "Comput. Softw. Big Sci.",
    volume = "7",
    number = "1",
    pages = "13",
    year = "2023"
}

@article{Qasim:2022rww,
    author = "Qasim, Shah Rukh and Chernyavskaya, Nadezda and Kieseler, Jan and Long, Kenneth and Viazlo, Oleksandr and Pierini, Maurizio and Nawaz, Raheel",
    title = "{End-to-end multi-particle reconstruction in high occupancy imaging calorimeters with graph neural networks}",
    eprint = "2204.01681",
    archivePrefix = "arXiv",
    primaryClass = "physics.ins-det",
    doi = "10.1140/epjc/s10052-022-10665-7",
    journal = "Eur. Phys. J. C",
    volume = "82",
    number = "8",
    pages = "753",
    year = "2022"
}

@article{Acosta:2023nuw,
    author = "Acosta, Fernando Torales and Karki, Bishnu and Karande, Piyush and Angerami, Aaron and Arratia, Miguel and Barish, Kenneth and Milton, Ryan and Mor{\'a}n, Sebasti{\'a}n and Nachman, Benjamin and Sinha, Anshuman",
    title = "{The optimal use of segmentation for sampling calorimeters}",
    eprint = "2310.04442",
    archivePrefix = "arXiv",
    primaryClass = "physics.ins-det",
    doi = "10.1088/1748-0221/19/06/P06002",
    journal = "JINST",
    volume = "19",
    number = "06",
    pages = "P06002",
    year = "2024"
}

@article{Alimena:2020web,
    author = "Alimena, Juliette and Iiyama, Yutaro and Kieseler, Jan",
    title = "{Fast convolutional neural networks for identifying long-lived particles in a high-granularity calorimeter}",
    eprint = "2004.10744",
    archivePrefix = "arXiv",
    primaryClass = "hep-ex",
    doi = "10.1088/1748-0221/15/12/P12006",
    journal = "JINST",
    volume = "15",
    number = "12",
    pages = "P12006",
    year = "2020"
}

@article{NA62:2023wzm,
    author = "Cortina Gil, Eduardo and others",
    collaboration = "NA62",
    title = "{Improved calorimetric particle identification in NA62 using machine learning techniques}",
    eprint = "2304.10580",
    archivePrefix = "arXiv",
    primaryClass = "hep-ex",
    reportNumber = "CERN-EP-2023-066",
    doi = "10.1007/JHEP11(2023)138",
    journal = "JHEP",
    volume = "11",
    pages = "138",
    year = "2023"
}

@article{Song:2023ceh,
    author = "Song, S. and Chen, J. and Liu, J. and Liu, Y. and Qi, B. and Shi, Y. and Wang, J. and Wang, Z. and Yang, H.",
    title = "{Study of residual artificial neural network for particle identification in the CEPC high-granularity calorimeter prototype}",
    eprint = "2310.09489",
    archivePrefix = "arXiv",
    primaryClass = "hep-ex",
    doi = "10.1088/1748-0221/19/04/P04033",
    journal = "JINST",
    volume = "19",
    number = "04",
    pages = "P04033",
    year = "2024"
}

@article{Charan:2023ldg,
    author = "Charan, Abtin Narimani",
    title = "{Particle identification with the Belle II calorimeter using machine learning}",
    eprint = "2301.11654",
    archivePrefix = "arXiv",
    primaryClass = "hep-ex",
    doi = "10.1088/1742-6596/2438/1/012111",
    journal = "J. Phys. Conf. Ser.",
    volume = "2438",
    number = "1",
    pages = "012111",
    year = "2023"
}

@article{Belayneh:2019vyx,
    author = "Belayneh, Dawit and others",
    title = "{Calorimetry with deep learning: particle simulation and reconstruction for collider physics}",
    eprint = "1912.06794",
    archivePrefix = "arXiv",
    primaryClass = "physics.ins-det",
    reportNumber = "FERMILAB-PUB-20-448-CMS",
    doi = "10.1140/epjc/s10052-020-8251-9",
    journal = "Eur. Phys. J. C",
    volume = "80",
    number = "7",
    pages = "688",
    year = "2020"
}

@article{Dimitrova:2025mbl,
    author = "Dimitrova, Kalina and Kozhuharov, Venelin and Nastaev, Ruslan and Petkov, Peicho",
    title = "{Cluster Reconstruction in Electromagnetic Calorimeters Using Machine Learning Methods}",
    eprint = "2505.24740",
    archivePrefix = "arXiv",
    primaryClass = "physics.ins-det",
    doi = "10.1088/1742-6596/3116/1/012004",
    journal = "J. Phys. Conf. Ser.",
    volume = "3116",
    number = "1",
    pages = "012004",
    year = "2025"
}

@article{CMSHGCAL:2024esz,
    author = "Aamir, M. and others",
    collaboration = "CMS HGCAL, CALICE AHCAL",
    title = "{Using graph neural networks to reconstruct charged pion showers in the CMS High Granularity Calorimeter}",
    eprint = "2406.11937",
    archivePrefix = "arXiv",
    primaryClass = "physics.ins-det",
    doi = "10.1088/1748-0221/19/11/P11025",
    journal = "JINST",
    volume = "19",
    number = "11",
    pages = "P11025",
    year = "2024"
}

@article{Akchurin:2024ffj,
    author = "Akchurin, Nural and others",
    title = "{Vertex Imaging Hadron Calorimetry Using AI/ML Tools}",
    eprint = "2408.15385",
    archivePrefix = "arXiv",
    primaryClass = "physics.ins-det",
    doi = "10.1051/epjconf/202532000026",
    journal = "EPJ Web Conf.",
    volume = "320",
    pages = "00026",
    year = "2025"
}

@article{Hashmani:2024ykk,
    author = {Hashmani, Raheem Karim and Akba{\c{s}}, Emre and Demirk{\"o}z, Melahat Bilge},
    title = "{A comparison of deep learning models for proton background rejection with the AMS electromagnetic calorimeter}",
    eprint = "2402.16285",
    archivePrefix = "arXiv",
    primaryClass = "hep-ex",
    doi = "10.1088/2632-2153/ad7cc0",
    journal = "Mach. Learn. Sci. Tech.",
    volume = "5",
    number = "4",
    pages = "045008",
    year = "2024"
}

@article{Polson:2021kvr,
    author = "Polson, L. and Kurchaninov, L. and Lefebvre, M.",
    title = "{Energy reconstruction in a liquid argon calorimeter cell using convolutional neural networks}",
    eprint = "2109.05124",
    archivePrefix = "arXiv",
    primaryClass = "physics.ins-det",
    doi = "10.1088/1748-0221/17/01/P01002",
    journal = "JINST",
    volume = "17",
    number = "01",
    pages = "P01002",
    year = "2022"
}

@article{Akchurin:2021afn,
    author = "Akchurin, N. and Cowden, C. and Damgov, J. and Hussain, A. and Kunori, S.",
    title = "{On the use of neural networks for energy reconstruction in high-granularity calorimeters}",
    eprint = "2107.10207",
    archivePrefix = "arXiv",
    primaryClass = "physics.ins-det",
    reportNumber = "APDL-2021-001",
    doi = "10.1088/1748-0221/16/12/P12036",
    journal = "JINST",
    volume = "16",
    number = "12",
    pages = "P12036",
    year = "2021"
}

@article{Simkina:2023ztj,
    author = {Simkina, Polina and Couderc, Fabrice and Malcl{\`e}s, Julie and Sahin, Mehmet {\"O}zg{\"u}r},
    title = "{Reconstruction of electromagnetic showers in calorimeters using Deep Learning}",
    eprint = "2311.17914",
    archivePrefix = "arXiv",
    primaryClass = "hep-ex",
    doi = "10.1140/epjc/s10052-024-12978-1",
    journal = "Eur. Phys. J. C",
    volume = "84",
    number = "6",
    pages = "639",
    year = "2024"
}

@article{Simsek:2024zhj,
    author = "Simsek, Ebru and Isildak, Bora and Dogru, Anil and Aydogan, Reyhan and Bayrak, Aydogan Burak and Ertekin, Seyda",
    title = "{CALPAGAN: Calorimetry for Particles Using Generative Adversarial Networks}",
    eprint = "2401.02248",
    archivePrefix = "arXiv",
    primaryClass = "hep-ex",
    doi = "10.1093/ptep/ptae106",
    journal = "PTEP",
    volume = "2024",
    number = "8",
    pages = "083C01",
    year = "2024"
}

@article{Dubinski:2023fsy,
    author = "Dubi{\'n}ski, Jan and Deja, Kamil and Wenzel, Sandro and Rokita, Przemys{\l}aw and Trzci{\'n}ski, Tomasz",
    title = "{Machine learning methods for simulating particle response in the zero degree calorimeter at the ALICE experiment, CERN}",
    eprint = "2306.13606",
    archivePrefix = "arXiv",
    primaryClass = "cs.CV",
    doi = "10.1063/5.0203567",
    journal = "AIP Conf. Proc.",
    volume = "3061",
    number = "1",
    pages = "040001",
    year = "2024"
}

@article{Rogachev:2022hjg,
    author = "Rogachev, Alexander and Ratnikov, Fedor",
    title = "{GAN with an Auxiliary Regressor for the Fast Simulation of the Electromagnetic Calorimeter Response}",
    eprint = "2207.06329",
    archivePrefix = "arXiv",
    primaryClass = "physics.data-an",
    doi = "10.1088/1742-6596/2438/1/012086",
    journal = "J. Phys. Conf. Ser.",
    volume = "2438",
    number = "1",
    pages = "012086",
    year = "2023"
}

@article{Bieringer:2022cbs,
    author = "Bieringer, Sebastian and Butter, Anja and Diefenbacher, Sascha and Eren, Engin and Gaede, Frank and Hundhausen, Daniel and Kasieczka, Gregor and Nachman, Benjamin and Plehn, Tilman and Trabs, Mathias",
    title = "{Calomplification {\textemdash} the power of generative calorimeter models}",
    eprint = "2202.07352",
    archivePrefix = "arXiv",
    primaryClass = "hep-ph",
    reportNumber = "DESY-22-031",
    doi = "10.1088/1748-0221/17/09/P09028",
    journal = "JINST",
    volume = "17",
    number = "09",
    pages = "P09028",
    year = "2022"
}

@article{Khattak:2021ndw,
    author = "Khattak, Gul Rukh and Vallecorsa, Sofia and Carminati, Federico and Khan, Gul Muhammad",
    title = "{Fast simulation of a high granularity calorimeter by generative adversarial networks}",
    eprint = "2109.07388",
    archivePrefix = "arXiv",
    primaryClass = "physics.ins-det",
    doi = "10.1140/epjc/s10052-022-10258-4",
    journal = "Eur. Phys. J. C",
    volume = "82",
    number = "4",
    pages = "386",
    year = "2022"
}

@article{Chekalina:2018hxi,
    author = "Chekalina, Viktoria and Orlova, Elena and Ratnikov, Fedor and Ulyanov, Dmitry and Ustyuzhanin, Andrey and Zakharov, Egor",
    editor = "Forti, A. and Betev, L. and Litmaath, M. and Smirnova, O. and Hristov, P.",
    title = "{Generative Models for Fast Calorimeter Simulation: the LHCb case}",
    eprint = "1812.01319",
    archivePrefix = "arXiv",
    primaryClass = "physics.data-an",
    doi = "10.1051/epjconf/201921402034",
    journal = "EPJ Web Conf.",
    volume = "214",
    pages = "02034",
    year = "2019"
}

@article{Paganini:2017dwg,
    author = "Paganini, Michela and de Oliveira, Luke and Nachman, Benjamin",
    title = "{CaloGAN : Simulating 3D high energy particle showers in multilayer electromagnetic calorimeters with generative adversarial networks}",
    eprint = "1712.10321",
    archivePrefix = "arXiv",
    primaryClass = "hep-ex",
    doi = "10.1103/PhysRevD.97.014021",
    journal = "Phys. Rev. D",
    volume = "97",
    number = "1",
    pages = "014021",
    year = "2018"
}

@article{Liu:2024kvv,
    author = "Liu, Qibin and Shimmin, Chase and Liu, Xiulong and Shlizerman, Eli and Li, Shu and Hsu, Shih-Chieh",
    title = "{Calo-VQ: Vector-Quantized Two-Stage Generative Model in Calorimeter Simulation}",
    eprint = "2405.06605",
    archivePrefix = "arXiv",
    primaryClass = "physics.ins-det",
    month = "may",
    journal = "arXiv preprint",
    year = "2024"
}

@article{Buhmann:2023bwk,
    author = {Buhmann, Erik and Diefenbacher, Sascha and Eren, Engin and Gaede, Frank and Kasieczka, Gregor and Korol, Anatolii and Korcari, William and Kr{\"u}ger, Katja and McKeown, Peter},
    title = "{CaloClouds: fast geometry-independent highly-granular calorimeter simulation}",
    eprint = "2305.04847",
    archivePrefix = "arXiv",
    primaryClass = "physics.ins-det",
    reportNumber = "DESY-23-061",
    doi = "10.1088/1748-0221/18/11/P11025",
    journal = "JINST",
    volume = "18",
    number = "11",
    pages = "P11025",
    year = "2023"
}

@article{Schnake:2024mip,
    author = {Schnake, Simon and Kr{\"u}cker, Dirk and Borras, Kerstin},
    title = "{CaloPointFlow II Generating Calorimeter Showers as Point Clouds}",
    eprint = "2403.15782",
    archivePrefix = "arXiv",
    primaryClass = "physics.ins-det",
    month = "3",
    journal = "arXiv preprint",
    year = "2024"
}

@article{Krause:2021ilc,
    author = "Krause, Claudius and Shih, David",
    title = "{Fast and accurate simulations of calorimeter showers with normalizing flows}",
    eprint = "2106.05285",
    archivePrefix = "arXiv",
    primaryClass = "physics.ins-det",
    doi = "10.1103/PhysRevD.107.113003",
    journal = "Phys. Rev. D",
    volume = "107",
    number = "11",
    pages = "113003",
    year = "2023"
}

@article{Kobylianskii:2024ijw,
    author = "Kobylianskii, Dmitrii and Soybelman, Nathalie and Dreyer, Etienne and Gross, Eilam",
    title = "{Graph-based diffusion model for fast shower generation in calorimeters with irregular geometry}",
    eprint = "2402.11575",
    archivePrefix = "arXiv",
    primaryClass = "hep-ex",
    doi = "10.1103/PhysRevD.110.072003",
    journal = "Phys. Rev. D",
    volume = "110",
    number = "7",
    pages = "072003",
    year = "2024"
}

@article{Favaro:2024rle,
    author = "Favaro, Luigi and Ore, Ayodele and Schweitzer, Sofia Palacios and Plehn, Tilman",
    title = "{CaloDREAM -- Detector Response Emulation via Attentive flow Matching}",
    eprint = "2405.09629",
    archivePrefix = "arXiv",
    primaryClass = "hep-ph",
    doi = "10.21468/SciPostPhys.18.3.088",
    journal = "SciPost Phys.",
    volume = "18",
    pages = "088",
    year = "2025"
}

@article{JUNO:2021vlw,
    author = "Abusleme, Angel and others",
    collaboration = "JUNO",
    title = "{JUNO physics and detector}",
    eprint = "2104.02565",
    archivePrefix = "arXiv",
    primaryClass = "hep-ex",
    doi = "10.1016/j.ppnp.2021.103927",
    journal = "Prog. Part. Nucl. Phys.",
    volume = "123",
    pages = "103927",
    year = "2022"
}

@article{LHAASO:2019qtb,
    author = "Addazi, Andrea and others",
    collaboration = "LHAASO",
    title = "{The Large High Altitude Air Shower Observatory (LHAASO) Science Book (2021 Edition)}",
    eprint = "1905.02773",
    archivePrefix = "arXiv",
    primaryClass = "astro-ph.HE",
    journal = "Chin. Phys. C",
    volume = "46",
    pages = "035001--035007",
    year = "2022"
}

@article{Sakharov:1967dj,
    author = "Sakharov, A. D.",
    title = "{Violation of CP Invariance, C asymmetry, and baryon asymmetry of the universe}",
    doi = "10.1070/PU1991v034n05ABEH002497",
    journal = "Pisma Zh. Eksp. Teor. Fiz.",
    volume = "5",
    pages = "32--35",
    year = "1967"
}

@article{CLAS:2003umf,
    author = "Mecking, B. A. and others",
    collaboration = "CLAS",
    title = "{The CEBAF Large Acceptance Spectrometer (CLAS)}",
    reportNumber = "JLAB-PHY-03-01",
    doi = "10.1016/S0168-9002(03)01001-5",
    journal = "Nucl. Instrum. Meth. A",
    volume = "503",
    pages = "513--553",
    year = "2003"
}

@article{Scoville:2006vq,
    author = "Scoville, Nick and others",
    title = "{The Cosmic Evolution Survey (COSMOS): Overview}",
    eprint = "astro-ph/0612305",
    archivePrefix = "arXiv",
    doi = "10.1086/516585",
    journal = "Astrophys. J. Suppl.",
    volume = "172",
    pages = "1--8",
    year = "2007"
}

@article{10.1111/j.1365-2966.2008.13535.x,
    author = {Da Cunha, Elisabete and Charlot, Stéphane and Elbaz, David},
    title = {A simple model to interpret the ultraviolet, optical and infrared emission from galaxies},
    journal = {Monthly Notices of the Royal Astronomical Society},
    volume = {388},
    number = {4},
    pages = {1595-1617},
    year = {2008},
    month = {08},
    issn = {0035-8711},
    doi = {10.1111/j.1365-2966.2008.13535.x},
    url = {https://doi.org/10.1111/j.1365-2966.2008.13535.x},
    eprint = {https://academic.oup.com/mnras/article-pdf/388/4/1595/3036508/mnras0388-1595.pdf},
}

@article{Gunderson:1979yi,
    author = "Gunderson, B. and Learned, J. and Mapp, J. and Reeder, D. D.",
    title = "{A Measurement of the Anti-neutron - Proton Cross-section as Low-energy}",
    reportNumber = "COO-088-024",
    doi = "10.1103/PhysRevD.23.587",
    journal = "Phys. Rev. D",
    volume = "23",
    pages = "587",
    year = "1981"
}

@article{BROOKHAVEN-HOUSTON-PENNSYLVANIASTATE-RICE:1987vhf,
    author = "Armstrong, T. and others",
    collaboration = "BROOKHAVEN-HOUSTON-PENNSYLVANIA STATE-RICE",
    title = "{Measurement of Anti-neutron Proton Total and Annihilation Cross-sections From 100-{MeV}/c to 500-{MeV}/c}",
    doi = "10.1103/PhysRevD.36.659",
    journal = "Phys. Rev. D",
    volume = "36",
    pages = "659--673",
    year = "1987"
}

@article{OBELIX:2000kga,
    author = "Iazzi, F. and others",
    collaboration = "OBELIX",
    title = "{Antineutron proton total cross-section from 50-MeV/c to 400-MeV/c}",
    doi = "10.1016/S0370-2693(00)00086-1",
    journal = "Phys. Lett. B",
    volume = "475",
    pages = "378--385",
    year = "2000"
}

@article{Belle:2000cnh,
    author = "Abashian, A. and others",
    collaboration = "Belle",
    title = "{The Belle Detector}",
    reportNumber = "KEK-PROGRESS-REPORT-2000-4",
    doi = "10.1016/S0168-9002(01)02013-7",
    journal = "Nucl. Instrum. Meth. A",
    volume = "479",
    pages = "117--232",
    year = "2002"
}

@article{Bossi:2008aa,
    author = "Bossi, F. and De Lucia, E. and Lee-Franzini, J. and Miscetti, S. and Palutan, M.",
    collaboration = "KLOE",
    title = "{Precision Kaon and Hadron Physics with KLOE}",
    eprint = "0811.1929",
    archivePrefix = "arXiv",
    primaryClass = "hep-ex",
    doi = "10.1393/ncr/i2008-10037-9",
    journal = "Riv. Nuovo Cim.",
    volume = "31",
    number = "10",
    pages = "531--623",
    year = "2008"
}

@article{Campbell:2016cea,
    author = "Campbell, Sarah",
    collaboration = "sPHENIX",
    title = "{sPHENIX: The next generation heavy ion detector at RHIC}",
    eprint = "1611.03003",
    archivePrefix = "arXiv",
    primaryClass = "physics.ins-det",
    doi = "10.1088/1742-6596/832/1/012012",
    journal = "J. Phys. Conf. Ser.",
    volume = "832",
    number = "1",
    pages = "012012",
    year = "2017"
}

@article{Belias:2023lkk,
    author = "Belias, A.",
    collaboration = "PANDA",
    title = "{Overview of the PANDA detector design at FAIR}",
    doi = "10.1142/S2010194523600017",
    journal = "Int. J. Mod. Phys. Conf. Ser.",
    volume = "51",
    pages = "2360001",
    year = "2023"
}

@article{Anderle:2021wcy,
    author = "Anderle, Daniele P. and others",
    title = "{Electron-ion collider in China}",
    eprint = "2102.09222",
    archivePrefix = "arXiv",
    primaryClass = "nucl-ex",
    reportNumber = "Frontiers of Physics, Volume 16 Issue (6):64701, 2021",
    doi = "10.1007/s11467-021-1062-0",
    journal = "Front. Phys. (Beijing)",
    volume = "16",
    number = "6",
    pages = "64701",
    year = "2021"
}

@article{CMS:2008xjf,
    author = "Chatrchyan, S. and others",
    collaboration = "CMS",
    title = "{The CMS Experiment at the CERN LHC}",
    doi = "10.1088/1748-0221/3/08/S08004",
    journal = "JINST",
    volume = "3",
    pages = "S08004",
    year = "2008"
}

@article{ATLAS:2008xda,
    author = "Aad, G. and others",
    collaboration = "ATLAS",
    title = "{The ATLAS Experiment at the CERN Large Hadron Collider}",
    doi = "10.1088/1748-0221/3/08/S08003",
    journal = "JINST",
    volume = "3",
    pages = "S08003",
    year = "2008"
}

@article{ALICE:2008ngc,
    author = "Aamodt, K. and others",
    collaboration = "ALICE",
    title = "{The ALICE experiment at the CERN LHC}",
    doi = "10.1088/1748-0221/3/08/S08002",
    journal = "JINST",
    volume = "3",
    pages = "S08002",
    year = "2008"
}

@article{LHCb:2008vvz,
    author = "Alves, Jr., A. Augusto and others",
    collaboration = "LHCb",
    title = "{The LHCb Detector at the LHC}",
    reportNumber = "LHCb-DP-2008-001",
    doi = "10.1088/1748-0221/3/08/S08005",
    journal = "JINST",
    volume = "3",
    pages = "S08005",
    year = "2008"
}

@article{CEPCStudyGroup:2018ghi,
    author = "Dong, Mingyi and others",
    collaboration = "CEPC Study Group",
    title = "{CEPC Conceptual Design Report: Volume 2 - Physics {\&} Detector}",
    eprint = "1811.10545",
    archivePrefix = "arXiv",
    primaryClass = "hep-ex",
    reportNumber = "IHEP-CEPC-DR-2018-02, IHEP-EP-2018-01, IHEP-TH-2018-01",
    month = "11",
    year = "2018",
    journal={arXiv preprint arXiv:1811.10545},
    doi={10.48550/arXiv.1811.10545},
}

@article{FCC:2018byv,
    author = "Abada, A. and others",
    collaboration = "FCC",
    title = "{FCC Physics Opportunities}: {Future Circular Collider Conceptual Design Report Volume 1}",
    reportNumber = "CERN-ACC-2018-0056",
    doi = "10.1140/epjc/s10052-019-6904-3",
    journal = "Eur. Phys. J. C",
    volume = "79",
    number = "6",
    pages = "474",
    year = "2019"
}

@misc{github,
  author       = {Yu, Hongtian and others},
  title        = {Mixed-representation Calorimetric Network},
  year         = {2025},
  howpublished = {\url{https://github.com/yuhongtian17/ViC}},
  note         = {GitHub repository}
}

\end{document}